\documentclass[aps,10pt,prd,notitlepage,nofootinbib]{revtex4-1}

\usepackage[utf8]{inputenc}
\usepackage{newtxtext,newtxmath}
\usepackage{bbold}
\usepackage{bm}
\usepackage{graphicx}
\usepackage[usenames,dvipsnames]{xcolor}
\usepackage{color}
\usepackage{tikz}
\usepackage[colorlinks=true,linkcolor=blue,urlcolor=blue,citecolor=blue]{hyperref}
\usepackage{amsmath,amssymb,amsfonts}
\usepackage{slashed}
\usepackage[english]{babel}
\usepackage{titlesec}
\usepackage{microtype}
\usepackage{dcolumn}
\usepackage{blindtext}
\usepackage{epsfig}
\usepackage{xcolor}
\usepackage{pifont}
\usepackage{dsfont,mathrsfs}
\usepackage{cancel}
\usepackage{bigints}

\usepackage{accents}
\usepackage{soul}
\usepackage{multirow}

\def\be {\begin{equation}}
\def\ee {\end{equation}}
\def\mn {{\mu\nu}}
\def\ba {\begin{eqnarray}}
\def\ea {\end{eqnarray}}
\def\nn {\nonumber}
\def\cm {{\cal M}}
\def\cl {{\cal L}}

\def\del {\partial}

\def\vp {\vec p}

\def\de {\delta}

\def\De {\Delta}

\newcommand{\FB}[1]{\left(#1\right)}
\newcommand{\SB}[1]{\left\{#1\right\}}
\newcommand{\TB}[1]{\left[#1\right]}

\begin{document}
\title{In-medium thermal conductivity and diffusion coefficients of a hot hadronic gas mixture}

\author{Utsab Gangopadhaya}
\email{utsabgango@vecc.gov.in}
\author{Snigdha Ghosh}
\email{snigdha.physics@gmail.com}
\author{Sourav Sarkar}
\email{sourav@vecc.gov.in}

\affiliation{Variable Energy Cyclotron Centre, 1/AF Bidhannagar, Kolkata 700 064, India}
\affiliation{Homi Bhabha National Institute, Training School Complex, Anushaktinagar, Mumbai - 400085, India}

\begin{abstract}
The relativistic kinetic theory approach has been employed to estimate for a hot mixture constituting of nucleons and pions, four transport coefficients that characterize heat flow and diffusion. Medium effects on the cross-section for binary collisions ($\pi\pi$,$\pi N$) have been taken into consideration by incorporating self-energy corrections to modify the  $\rho$ and $\sigma$ meson propagators for $\pi\pi$ interaction and the $\Delta$ propagator for $\pi N$ interaction. The temperature dependence of the various coefficients have been studied for several values of the baryon chemical potential. 

\end{abstract}

\maketitle

\section{INTRODUCTION}

Ultra relativistic heavy ion collision provides us with the unique possibility to investigate strong interacting matter in a deconfined state with coloured degrees of freedom, known as quark gluon plasma(QGP)~\cite{Book}. Production and investigation of such matter has been one of the primary goals of the experiments being pursued at the RHIC at BNL and LHC at CERN. Relativistic hydrodynamics is an effective tool which has been widely used to investigate the evolution of the properties of the matter created during these collisions. This is because soon after the collision the system is believed to reach a state where the fluctuation of the thermodynamic quantities describing the system is much smaller compared to the value of the thermodynamic quantities themselves. One of the turning points in these studies has been the observation of a large elliptic flow in 200 AGeV Au+Au collisions at RHIC which led to the conclusion that the system behaves like a near perfect fluid with a small but finite value of shear viscosity to entropy density ratio $(\eta/s)$ (see e.g.~\cite{Luzum}).  Further investigations have revealed that whereas $\eta/s$ shows a minimum near the QGP-hadron gas crossover~\cite{Csernai} the bulk viscosity to entropy density ratio $\zeta/s$ shows large values~\cite{Karsch}.

The fact that transport coefficients not only serve to characterize the nature of matter produced but are useful as signatures of phase transition prompted a lot of activity in recent times both above and below the QGP-hadron gas crossover temperature $T_c$. These coefficients go as inputs to the dissipative hydrodynamic equations for which the second order formulation of Israel and Stewart~\cite{Israel} is currently most widely used. They are calculated using either linear response theory where they appear in the form of two-point functions or in the kinetic theory approach which is more widely used owing to better computational efficiency~\cite{Jeon}. For QCD matter transport coefficients have been obtained using linear response as in Refs.~\cite{Defu,Moore} and Refs.~\cite{Thoma,Baym,Kajantie,Amy} using kinetic theory.

Quite a substantial amount of recent literature also deals with transport coefficients of hot hadronic matter. We refer to some of them in the following.  Shear viscosity has been derived using a hadron resonance gas model in Ref.~\cite{Noronha, Wiranata, Pal, Kadam}. In the Kinetic theory approach viscosities have been estimated using parametrized cross section extracted from empirical data in Refs. ~\cite{Gavin, Itakura, Prakash, Davesne}, or using lowest order chiral perturbation theory  in Refs.~\cite{Dobado, Chen}.
NJL model has been used in kinetic theory  to predict viscous coefficients in Ref.~\cite{Heckmann}. Again, chiral perturbation theory has been used in Ref.~\cite{Lu} to calculate the bulk viscosity taking into consideration the number changing inelastic processes, and linear sigma model has been used to study the behavior of bulk viscosity around phase transition in Ref.~\cite{Dobado2}. 

Compared to the viscosities, the thermal conductivity and diffusion coefficients have received much less attention. This may be due to the absence of a conserved quantum number, the baryon number being insignificantly small for systems produced at RHIC and LHC. However at FAIR energies~\cite{CBMBook} or in the Beam Energy Scan (BES) program at RHIC, the baryon chemical potential is expected to be significant and baryon number will play a significant role in determining the thermal and diffusion coefficients. A careful analysis also reveals that a system in which the total number of particle is conserved can also sustain thermal conduction or diffusion if the number of particles of individual species is also conserved. Such a scenario is reached as the system expands and cools, it reaches chemical freeze out as the collisions become mostly elastic.

Diffusion coefficients arise in the treatment of a multicomponent gas in addition to thermal conductivity~\cite{DegrootBook}. For the pion-nucleon system under study there arise two thermal coefficients (thermal conductivity and Dufour coefficient) and two diffusion coefficients (diffusion and thermal diffusion coefficient). We have not encountered an elaborate study of the temperature and density dependence of these coefficients in the literature. Additionally we have introduced medium effects in the dynamics to obtain a more realistic description. Recall that in the kinetic theory approach the dynamic input appears in the collision integral as the cross section. In most of the analyses employing kinetic theory to evaluate transport coefficients vacuum cross section was employed. Since the system under study is produced during the later stages of heavy ion collisions and is presumably at a high temperature and/or density we have incorporated medium effects in the cross-section and this is a novel feature in our approach. For the case of a pion gas the medium effects on the $\pi\pi$ cross-section were incorporated by means of self-energy corrections on the intermediate $\rho$ and $\sigma$ meson propagators appearing in the scattering matrix elements. A significant modification of the cross-section due to medium effects resulted in a noticeable consequences on the temperature dependence of the viscosities~\cite{Sukanya1}, thermal conductivity~\cite{Sukanya2} and the relaxation time of flows~\cite{Sukanya3} for a hot pion gas. For the case of a hadronic gas mixture of nucleons and pions the in-medium $\pi N$ cross-section obtained using a modified $\Delta$ propagator~\cite{Snigdha} resulted in significant changes in the viscous coefficients~\cite{Utsab}.
 
In the present work we obtain the temperature and (baryon) density dependence of the thermal and diffusion coefficients of a hadronic gas mixture of pions and nucleons introducing in-medium cross-sections in the kinetic theory approach using the relaxation time approximation. In the next section the expressions for the coefficients are discussed followed by a section on the evaluation of in-medium cross-section using thermal field theoretic methods. Results are followed by appendices containing calculational details.

\section{Formalism}
We begin with a brief description of the hydrodynamic equations. The energy momentum tensor $T^{\mu\nu}$ can be decomposed 
into reversible $T^{(0)\mu\nu}$, and irreversible  $T^{(1)\mu\nu}$ contributions, given by
\begin{eqnarray}
 T^{(0)\mu\nu} &=& enU^{\mu}U^{\nu}-P\Delta^{\mu\nu} \nonumber\\
T^{(1)\mu\nu} &=& \left[ (I^{\mu}_{q}+h\Delta^{\mu\sigma}N_{\sigma})U^{\nu}+(I^{\nu}_{q}+h\Delta^{\nu\sigma}N_{\sigma})U^{\mu}\right]+\Pi^{\mu\nu}~,
\label{mc0}
\end{eqnarray}

 where, $P$, $e$ and $n$ are the pressure, energy density per particle and, particle density respectively;  $U^{\mu}$ is the hydrodynamic four velocity and $\Delta^{\mu\nu}=g^{\mu\nu}-U^{\mu}U^{\nu}$, is the projection operator. 
 In the above equations, $I^{\mu}_{q}$ and $N^{\mu}$ are the heat flow and the total particle four-flow density respectively, $\Pi^{\mu\nu}$ contains the shear and the bulk part of the irreversible momentum flow. 
Throughout this paper the convention of metric that has been used is $g^{\mu\nu}=diag(1,-1,-1,-1)$.

It follows from from Eq.~(\ref{mc0}) that the heat transfer can be expressed as
\begin{eqnarray}
  I^{\mu}_{q} =T^{(1)\mu\nu}U_{\nu}-h\Delta^{\mu\sigma}N_{\sigma}~,
\end{eqnarray}
since, $U^{\mu}U_{\mu}=1$ and $\Pi^{\mu\nu}U_{\nu}=0$. 
In linear hydrodynamic theory that satisfies the second law of thermodynamics, the relation 
between the thermodynamic forces and the reduced heat flow and particle flow for a binary mixture at mechanical equilibrium (i.e. $\bigtriangledown^{\mu}P=0$) is given by~\cite{Degroot1,Degroot2},
\begin{eqnarray}
\bar{I}^{\mu} &=& I^{\mu}-\sum_{k=1}^{2}h_{k}N_{k}^{\mu} \nonumber\\
\bar{I}^{\mu} &=& \lambda \bigtriangledown^{\mu}T+D^{'}_{T}n x_{1}T \FB{\frac{\partial \mu_{1}}{\partial x_{1}}}_{P,T}\bigtriangledown^{\mu}x_{1} 
\label{he1}\\
I^{\mu}_{k} &=& N^{\mu}_{k}-x_{k}N^{\mu} \nonumber\\
I^{\mu}_{k} &=& D_{F}n\bigtriangledown^{\mu}x_{1}+D_{T}nx_{1}x_{2}\bigtriangledown^{\mu}T ~,
\label{de1}
\end{eqnarray}
In the above equations, $x_1=n_1/n$, $x_2=n_2/n$ are the the concentration of first and the second species respectively; $n_1$ and $n_2$ being the particle density of the individual species. Thermal conductivity $\lambda$ is the coefficient that appears in the Fourier's law of heat conduction, which describes that transfer of energy due to temperature gradient alone~\cite{Demirel}. Dufour coefficient $D'_{T}$ on the other hand parametrises the Dufour effect which represents energy flow due to diffusion of one component under isothermal condition, or in other words due to the concentration gradient of one of the component. The Diffusion coefficient $D_F$ is the coefficient that parametrises the flow of a particular component due to its concentration gradient. The flow of a particular component due to temperature gradient alone is known as Soret's effect, and the coefficient that parametrise it is known as thermal diffusion coefficient $D_{T}$.  

 In the transport theory approach the transfer of momentum and energy is due to collisions between the constituent particles. The diffusion flow and the irreversible part of the energy momentum tensor is non-trivial when the system is not in local equilibrium. For slight deviation from equilibrium the local distribution function can be expressed as,
 \begin{eqnarray}
 f_{k}(x,p)=f^{(0)}_{k}(x,p)+\de f_{k}(x,p)~~,~~~~\de f_{k}(x,p)=f_{k}^{(0)}(x,p)\TB{1\pm f^{(0)}_{k}(x,p)}\phi_{k}(x,p)~.
 \label{ff}
 \end{eqnarray}
  Here $f^{(0)}_{k}$ is the equilibrium (Bose-Einstein or Fermi-Dirac) distribution function and $\delta f_{k}$, represents the slight deviation from equilibrium of the $k^{th}$ particle species. The $\pm$ sign in the expression of $\delta f_{k}$, denotes the Bose enhancement or the Pauli blocking. The quantity $\phi_{k}$ parametrizes the deviation. Thus the heat flow and the diffusion flow is given by,
  \begin{eqnarray}
  \bar{I}^{\mu}_{q} &=& \sum_{k=1}^{2}\int d\Gamma_{k}  \FB{p_{k}^\nu U_\nu-h_k} p_{k}^\mu \delta f_{k}.
  \label{he2}\\
  I^{\mu}_{1} &=& \sum_{k=1}^{2}\int d\Gamma_{k} \FB{\delta_{1k}-x_k}p_{k}^{\mu}\delta f_{k}
   \label{de2}  
  \end{eqnarray}
  where $d\Gamma_k=d^3p_k/(2\pi)^3E_k$ and, $E_k=\sqrt{\vec{p}_k^2+m_k^2}$. On the right side of Eq.(\ref{he1}) and Eq.(\ref{de1}) we find space derivative of temperature and the concentration of 1st species. The right hand side of Eq. (\ref{he2}) and Eq.(\ref{de2}) which refers to the same quantity as expressed in Eq.(\ref{he1}) and Eq.(\ref{de1}) involves  integral over the particle three momentum of both the species. In order that these two set of equations conform with each other, $\phi_{k}$ must be a linear combination of the space derivative of temperature and the concentration of first species with proper coefficients~\cite{Utsab},
  \begin{eqnarray}
  \phi_{k}=-B^{(k)q}_{\mu}\frac{\bigtriangledown^{\mu}T}{T}-B^{(k1)}_{\mu}\frac{1}{x_2}\FB{\frac{\partial \mu_1}{\partial x_1}}_{P,T}\bigtriangledown ^{\mu}x_1 ~.
  \label{phi}
  \end{eqnarray}
Substituting the above expression for $\phi_{k}$ in  Eq.(\ref{he2}) and Eq.(\ref{de2}), with the help of Eq.(\ref{ff}) and then comparing the coefficients of space derivate of temperature and concentration of the first species we get, 
\begin{eqnarray}
\lambda&=&\FB{\frac{L_{qq}}{T}} 
\label{eq.lambda.0}\\
D_{T}^{'}&=&\FB{\frac{L_{q1}}{nx_2x_1T}} 
\label{Dufour0}\\
D_F&=&\FB{\frac{L_{11}}{nx_2}}\FB{\frac{\partial \mu_1}{\partial x_1}}_{P,T} 
\label{Diffusion0}\\
D_{T}&=&\FB{\frac{L_{1q}}{nx_2x_1T}}~,
\label{TDif0}
\end{eqnarray}
where,
\begin{eqnarray}
L_{qq}&=&-\frac{1}{3}\sum_{k=1}^{2}\int d\Gamma_{k}(p_{k}\cdot U-h_{k})p_{k}^{\sigma}\Delta^{\alpha}_{\sigma}B^{(k)q}_{\alpha}f_{k}^{(0)}[1\pm f^{(0)}_{k}]
\label{lambda}\\
L_{q1}&=&-\frac{1}{3}\sum_{k=1}^{2}\int d\Gamma_{k}(p_{k}\cdot U-h_{k})p_{k}^{\sigma}\Delta^{\alpha}_{\sigma}B^{(k1)}_{\alpha}f_{k}^{(0)}[1\pm f^{(0)}_{k}]
\label{Dufour}\\
L_{11}&=&-\frac{1}{3}\sum_{k=1}^{2}\int d\Gamma_{k}(\delta_{k1}-x_{k})p_{k}^{\sigma}\Delta^{\alpha}_{\sigma}B^{(k1)}_{\alpha}f_{k}^{(0)}[1\pm f^{(0)}_{k}]
\label{Diffusion}\\
L_{1q} &=& -\frac{1}{3}\sum_{k=1}^{2}\int d\Gamma_{k}(\delta_{k1}-x_{k})p_{k}^{\sigma}\Delta^{\alpha}_{\sigma}B^{(k)q}_{\alpha}f_{k}^{(0)}[1\pm f^{(0)}_{k}]~.
\label{TDif}
\end{eqnarray}
To get the different coefficients we need to find $B^{(1)q}_{\alpha}$, $B^{(2)q}_{\alpha}$, $B^{(11)}_{\alpha}$ and $B^{(21)}_{\alpha}$, which can be obtained by solving the relativistic transport equation for two particle system.
\begin{eqnarray}
p_{k}^\mu\partial_\mu f_{k}(x,p)=\sum_{l=1}^{2}\FB{\frac{g_{l}}{1+\delta_{kl}}}C_{kl}[f_k],
\label{treq} 
\end{eqnarray}
 
  $g_{l}$ being the degeneracy of $l^{th}$ species of particle and the collision integral for binary elastic collisions $p_{k}+p_{l}\rightarrow p^{'}_{k}+p_{l}^{'}$ is given by,
\begin{eqnarray}
 C_{kl}[f]&=&\int\int\int d\Gamma_{p_{l}} d\Gamma_{p'_{k}} d\Gamma_{p'_{l}}
 \left[f_{k}\FB{x,p'_{k}}f_{l}\FB{x,p_{l}'} \SB{1+f_{k}\FB{x,p_{k}}}
 \SB{1+f_{l}\FB{x,p_{l}}} \right.\nonumber\\
 && \left.-~f_{k}\FB{x,p_{k}}f_{l}\FB{x,p_{l}}\SB{1+f_{k}\FB{x,p'_{k}}}\SB{1+f_{l}\FB{x,p'_{l}}}\right]\ W_{kl} ~,
 \label{coll}
\end{eqnarray}
 with the interaction rate $
 W_{kl}=\FB{\frac{s}{2}}\FB{\frac{d\sigma_{kl}}{d\Omega}}(2\pi)^6\delta^4(p_{k}+p_{l}-p'_{k}-p_{l}')$,
  where $s=(p_k+p_l)^2$ is the Mandelstam variable.
 The collision term on the right hand side of Eq.(\ref{treq}) in the relaxation time approximation (RTA)~\cite{Utsab,Prakash}
 is expressed as the deviation of the distribution function over the thermal relaxation time 
$\tau_k$ which is actually a measure of the time scale for restoration of the out of equilibrium distribution
to its local equilibrium value. we thus have 
 \begin{eqnarray}
 \sum_{l=1}^{2}\FB{\frac{g_{l}}{1+\delta_{kl}}}C_{kl}[f_k]\simeq-\FB{\frac{\delta f_k}{\tau_k}}=
 -\TB{\frac{f_k -f_{k}^{(0)}}{\tau_k}}~.
 \label{relax} 
 \end{eqnarray}
The relaxation time $\tau_{k}$ is taken as the inverse
of the reaction rate of the $k^{th}$ particle the explicit expression for which will be given in the next section. Note that there
are other ways of obtaining the relaxation time. Transport relaxation rates can be used besides other possible parametrizations. Moreover, though
RTA offers a simple and reasonably accurate way to handle the collision kernel, for better precision the 9-dimensional
collision integral given by Eq.~(\ref{coll}) needs to be considered along with more advanced methods of simplification.
 
 Expanding the distribution function using Chapman $(f_{k}=f_{k}^{(0)}+\epsilon f_{k}^{(1)}+\epsilon ^{2}f_{k}^{(2)}+...)$ expansion in terms of the Knudsen number, and restricting ourself to the first order, the transport equation assumes the form~\cite{Utsab,Degroot2}. 

 \begin{eqnarray} 
 p^{\mu}_{k}U_{\mu}Df^{(0)}_{k}+p^{\mu}_{k}\bigtriangledown _{\mu}f^{(0)}_{k}=-\FB{\frac{ f_k^{(1)}}{\tau_k}}~,
 \label{teq2}
 \end{eqnarray}
 where, $D=U^{\nu}\partial_{\nu}$, $\bigtriangledown _{\mu}=\bigtriangleup_{\mu \nu}\partial^{\nu}$ and $f_{k}^{(1)}=\delta f_k$. 
The covariant derivative on the left hand side has been split into a time like part and a space like part. 
Using the conservation equation the time derivative of the different thermodynamic parameters are turned into space derivatives of temperature and fluid velocity, the left hand side of Eq.(\ref{teq2}) becomes, 
 \begin{eqnarray}
 \frac{f_{k}^{(0)}(1\pm f_{k}^{(0)})}{TE_k}\left[Q_{k}\FB{\partial_{\nu}u^{\nu}}-
 \langle p_{k}^{\mu}p_{k}^{\nu} \rangle \langle \del_{\mu} u_{\nu}
 \rangle+(p_k^{\sigma}U_{\sigma})p_k^{\mu}\FB{\frac{\bigtriangledown_{\mu}T}{T}-\frac{\bigtriangledown_{\mu}P}{nh}}+Tp_k^{\mu}\bigtriangledown_{\mu}\FB{\frac{\mu_k}{T}}\right] ~.
 \label{teq2left}
 \end{eqnarray}
 Here $Q_{k}=T^2\TB{-\frac{1}{3}z_{k}^2+\tau_{k}^2\FB{\frac{4}{3}-\gamma}+\tau_{k}
 \SB{\FB{\gamma_{k}''-1}\hat{h}_{k}-~\gamma_{k}'''}}$, where
 $z_{k}=\FB{\frac{m_k}{T}}$, $\tau_{k}=\FB{\frac{p_{k}\cdot U}{T}}$ and $\hat{h}_{k}=\FB{\frac{h_k}{T}}$
 with $h_k$ is the enthalpy per particle belonging to $k^{th}$ species. The 
 details of the calculation along with the expressions of
 $\gamma$'s and $z$'s will be given in Appendix-\ref{sec.appendix.a}. Since we intend to extract the thermal and diffusion coefficients, we neglect the terms including the space derivative of the hydrodynamic velocity. Using the Gibbs Duhem relation and Eq.~(\ref{teq2left}), Eq.~(\ref{teq2}) can be written for a two component system as 
 \begin{eqnarray} 
 \frac{f_{k}^{(0)}(1+ f_{k}^{(0)})}{TE_k}\left[\SB{p_k \cdot U-h+(\delta_{k1}-x_1)
 T^2\FB{\frac{\partial}{\partial T}\FB{\frac{\mu_1}{T}}_{P,x_1}-\frac{\partial}{\partial T}\FB{\frac{\mu_{2}}{T}}_{P,x_1}}}p_k^{\mu}
\FB{\frac{\bigtriangledown_{\mu}T}{T}} \right.\nonumber\\ \left.
 +\frac{(\delta_{k1}-x_1)}{x_{2}}\FB{\frac{\partial \mu_1}{\partial x_1}}_{P,T}p_k^{\mu}\bigtriangledown_{\mu}x_1\right]=-\FB{\frac{\delta f_k}{\tau_k}}~.
 \label{teq3}
 \end{eqnarray}
 The details of the above calculation is given in Appendix-\ref{sec.appendix.b}. Substituting the expression for $\delta f_{k}$ from Eq.(\ref{ff}) in Eq.(\ref{teq3}) using the expression for $\phi_{k}$ from Eq.(\ref{phi}) and comparing the coefficients of the thermodynamic forces we get.
\begin{eqnarray}
B^{(k)q}_{\nu}\bigtriangleup^{\mu\nu}&=&\frac{\tau_{k}}{E_kT}\left[p_k\cdot U-h+(\delta_{k1}-x_1)T^2\FB{\frac{\partial}{\partial T}\FB{\frac{\mu_1}{T}}_{P,x_1}-\frac{\partial}{\partial T}\FB{\frac{\mu_{2}}{T}}_{P,x_1}}\right]\bigtriangleup^{\mu\nu}p_{k\nu}\\
B^{(k1)}_{\nu}\bigtriangleup^{\mu\nu}&=&\frac{\tau_{k}}{E_kT}(\delta_{k1}-x_1)\bigtriangleup^{\mu\nu}p_{k\nu}
\end{eqnarray}
 Using the above expressions in Eq.(\ref{lambda}) to Eq.(\ref{TDif}) we get,
 \begin{eqnarray} 
 L_{qq}&=&\frac{1}{3T} \sum_{k=1}^{2} g_k\int\frac{d^3p_k}{\FB{2\pi}^3}\FB{\frac{\vec{p}_k^2}{E_k^2}} (p_k\cdot U-h_k)\Big[p_k.U-h+(\delta_{k1}-x_1)T^2\beta\Big] \tau_k f_k^{(0)}(1\pm f_k^{(0)}) 
 \label{eq.lambda.1}\\
 L_{q1}&=&\frac{1}{3T} \sum_{k=1}^{2} g_k\int\frac{d^3p_k}{\FB{2\pi}^3}\FB{\frac{\vec{p}_k^2}{E_k^2}} (p_k\cdot U-h_k)(\delta_{k1}-x_1) \tau_k f_k^{(0)}(1\pm f_k^{(0)})\\
 L_{11}&=&\frac{1}{3T} \sum_{k=1}^{2} g_k\int\frac{d^3p_k}{\FB{2\pi}^3}\FB{\frac{\vec{p}_k^2}{E_k^2}} (\delta_{k1}-x_1)^2 \tau_k f_k^{(0)}(1\pm f_k^{(0)})\\
 L_{1q}&=&\frac{1}{3T} \sum_{k=1}^{2} g_k\int\frac{d^3p_k}{\FB{2\pi}^3}\FB{\frac{\vec{p}_k^2}{E_k^2}} \Big[p_k\cdot U-h+(\delta_{k1}-x_1)T^2\beta\Big](\delta_{k1}-x_1) \tau_k f_k^{(0)}(1\pm f_k^{(0)})
 \end{eqnarray}
 %
 %
%
where $\beta=\frac{\partial}{\partial T}\FB{\frac{\mu_1}{T}}_{P,x_1}-\frac{\partial}{\partial T}\FB{\frac{\mu_{2}}{T}}_{P,x_1}$.The expressions for $\frac{\partial}{\partial T}\FB{\frac{\mu_1}{T}}_{P,x_1}$, $\frac{\partial}{\partial T}\FB{\frac{\mu_2}{T}}_{P,x_1}$ and $\FB{\frac{\partial \mu_1}{\partial x_1}}_{P,T}$  
 needed to calculate the transport coefficients has been derived in Appendix-\ref{sec.appendix.c}. 


\section{Dynamical Inputs}
We now focus on the calculation of the relaxation times. Inter-particle collisions are responsible for transport 
phenomena and so dynamical information in the transport equation enter through the scattering cross section. For the case 
at hand spices 1 \& 2 denote the nucleon and the pion respectively. The relaxation times for nucleons and pions are coupled and involve the $N\pi$, 
$\pi\pi$ and $NN$ cross sections.They are respectively given by
\begin{eqnarray}
\tau_N^{-1}&=&\tau_{N\pi}^{-1}+\tau_{N N}^{-1}\nn\\
\tau_\pi^{-1}&=&\tau_{\pi\pi}^{-1}+\tau_{\pi N}^{-1}
\label{rel_piN}
\end{eqnarray}
where
\begin{eqnarray}
\tau_{kl}^{-1}(p_k) &=& \FB{\frac{\nu_{l}}{1+\de_{kl}}} \frac{1}{2E_{k}} \int\int\int d\Gamma_{p_{l}} d\Gamma_{p'_{k}} d\Gamma_{p'_{l}} \FB{2\pi}^4
\delta^{4}(p_{k}+p_{l}-p'_{k}-p'_{l}) \nn \\
&& \times~|\mathcal{M}_{k+l\rightarrow k+l}|^{2} \times \frac{f_{l}^{0} (1\pm f_{k}^{'0}) (1\pm f_{l}^{'0})}{(1\pm f_{k}^{0})}~, 
\label{eq-EC26}
\end{eqnarray}
 $|\mathcal{M}_{k+l\rightarrow k+l}|$ being the amplitude for binary 
elastic scattering processes. A popular approach is to use phenomenological amplitudes designed to reproduce experimental 
data of elastic scatterings. However, the system under consideration is produced in the later stage of HICs and in presumably at 
at high temperature and/or baryon density where the vacuum amplitude could be modified by many body effects. A realistic treatment 
thus suggests the use of in-medium amplitudes. Here we adopt a dynamical approach where standard effective interactions are 
used to evaluate scattering amplitudes. Considering resonant scattering in the s-channel the propagation in this channel 
are replaced by effective ones obtained by a Dyson-Schwinger sum of one-loop self energy diagrams. 

Let us first consider elastic $\pi N$ scattering. Considering $\mathcal{L}_{\pi N\Delta} = \frac{f_{\pi N\Delta}}{m_\pi}\bar{\De}^\mu\vec T^\dagger\partial_\mu\vec{\pi}\psi + H.c.$ with $f_{\pi N\Delta}=2.8$, the scattering proceeds 
via exchange of the $\Delta$ baryon. The propagation of the $\Delta$ is modified in the hot/dense medium. This is quantified 
through the self energy evaluated using standard methods of thermal field theory. The spin averaged expressions for 
the real and imaginary parts of $\Delta$ self energy are given by,
\begin{eqnarray}
\text{Re}~\Pi_\Delta(q) &=& \sum_{ i \in \{\pi,\rho\}}~ \sum_{j \in \{N,\Delta\}} 
\int\frac{d^3k_i}{(2\pi)^3} \frac{1}{2 \omega_{k_i}\omega_{p_j}}\mathcal{P}\left[
\left(\frac{n_+^{k_i}\omega_{p_j} \mathcal{N}_{ij}^\Delta(k_i^0=\omega_{k_i})}{(q_0-\omega_{k_i})^2-\omega_{p_j}^2}\right) +
\left(\frac{n_-^{k_i}\omega_{p_j} \mathcal{N}_{ij}^\Delta(k_i^0=-\omega_{k_i})}{(q_0+\omega_{k_i})^2-\omega_{p_j}^2}\right)-\right. \nn\\ 
&&\hspace{3cm}\left.\left(\frac{\tilde{n}_+^{p_j}\omega_{k_i} \mathcal{N}_{ij}^\Delta (k_i^0=q_0-\omega_{p_j})}{(q_0-\omega_{p_j})^2-\omega_{k_i}^2}\right) -
\left(\frac{\tilde{n}_-^{p_j}\omega_{k_i} \mathcal{N}_{ij}^\Delta (k_i^0=q_0+\omega_{p_j})}{(q_0+\omega_{p_j})^2-\omega_{k_i}^2}\right) \right] \label{eq.repi.delta}
\end{eqnarray}
and
\begin{eqnarray}
\text{Im}~\Pi_\Delta(q)&=& -\pi\epsilon(q_0) \sum_{ i \in \{\pi,\rho\}} ~ \sum_{j \in \{N,\Delta\}} 
\int\frac{d^3k_i}{(2\pi)^3}\frac{1}{4 \omega_{k_i}\omega_{p_j}}\times\nn \\
&&\left[\frac{}{}\mathcal{N}_{ij}^\Delta(k_i^0=\omega_{k_i})\left\{(1+n_+^{k_i}-\tilde{n}_+^{p_j})
\delta(q_0-\omega_{k_i}-\omega_{p_j})+(-n_+^{k_i}-\tilde{n}_-^{p_j})
\delta(q_0-\omega_{k_i}+\omega_{p_j})\right\} \right. \nn\\ 
&& \left.+ \mathcal{N}_{ij}^\Delta(k_i^0=-\omega_{k_i})\left\{(-1-n_-^{k_i}+\tilde{n}_-^{p_j})
\delta(q_0+\omega_{k_i}+\omega_{p_j})+
(n_-^{k_i}+\tilde{n}_+^{p_j})\delta(q_0+\omega_{k_i}-\omega_{p_j})\right\}\frac{}{}\right]~.
\label{eq.impi.delta}
\end{eqnarray}
In Eqs.~(\ref{eq.repi.delta}) and (\ref{eq.impi.delta}), the distribution functions for mesons and baryons are given by 
$n_\pm^{k_i} = \left[ e^{\beta\left(\omega_{k_i}\mp \mu_i \right)}-1 \right]^{-1}$ and 
$\tilde{n}_\pm^{p_j} = \left[ e^{\beta\left(\omega_{p_j}\mp \mu_j \right)}+1 \right]^{-1}$ respectively with 
$\omega_{k_i} = \sqrt{\vec{k}_i^2+m_i^2}$.
The expressions for $N_{ij}^\Delta$ may be found in Ref~\cite{Snigdha}. 
For $\{i,j\}\equiv\{\pi,\Delta\}$ and $\{\rho,N\}$, the 
contribution to the self energy have been further folded with the vacuum spectral function of $\Delta$ and $\rho$ 
respectively , details of which can be found in Ref.~\cite{Snigdha}. The real part shifts the pole at best by a small amount 
and the imaginary part modifies the width. The first term in Eq.~(\ref{eq.impi.delta}) is the contribution from decay and 
formation of $\Delta$ baryon weighted by thermal factors; the fourth term is due to scattering processes in the medium 
leads to the absorption of the $\Delta$; the second and third term do not contribute for the physical time-like momenta 
of $\Delta$ defined in terms of $q^0 > 0$ and $q^2 > 0$. 
\begin{figure}
\centering
\includegraphics[scale=0.3,angle=-90]{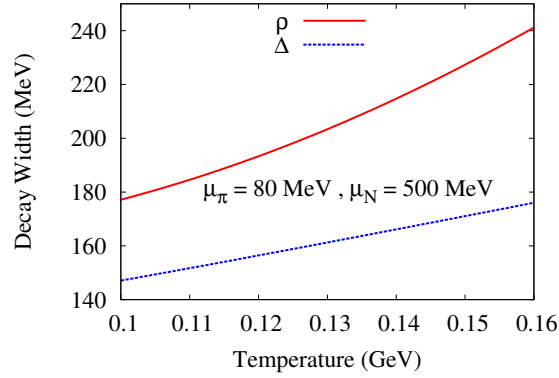}
\caption{Decay width of $\rho$ meson and $\Delta$ baryon as a function of temperature at constant pion and 
	nucleon chemical potential of 80 MeV and 500 MeV respectively.}
\label{fig:decay_width}
\end{figure}
The decay width of $\Delta$ is defined as,
\begin{equation}
\Gamma_\Delta \left( T,\mu_\pi, \mu_N \right) = - \frac{\text{Im}~\Pi_\Delta(q^0=m_\Delta,\vec{q}=\vec{0})}{m_\Delta}~.
\end{equation}

Note that in heavy ion collisions 
pions get out of chemical equilibrium
at $T\sim$ 170 MeV and a corresponding chemical potential starts building up
with decrease in temperature. The kinetics of the gas is then dominated by 
elastic collisions. We take the temperature dependent pion chemical potential from 
Ref.~\cite{Hirano} 
which reproduces the slope of the transverse momentum spectra of identified
hadrons observed in experiments.
Here, by
fixing the ratio of entropy and number density $s/n$ to the value at chemical freeze-out where $\mu_\pi=0$,
one can go down in
temperature up to the kinetic freeze-out by increasing the pion chemical
potential. This provides the temperature dependence leading to $\mu_\pi(T)$
whose value starts from zero at chemical freezeout and rises to a maximum at kinetic freezeout. The temperature dependence is parametrized as
\be
\mu_\pi(T)=a+bT+cT^2+dT^3 \nn
\ee
with $a=0.824$, $b=3.04$, $c=-0.028$, $d=6.05\times 10^{-5}$ and $T$, $\mu_\pi$ in MeV.  

\begin{figure}[h]
	\centering
	\includegraphics[scale=0.3,angle=-90]{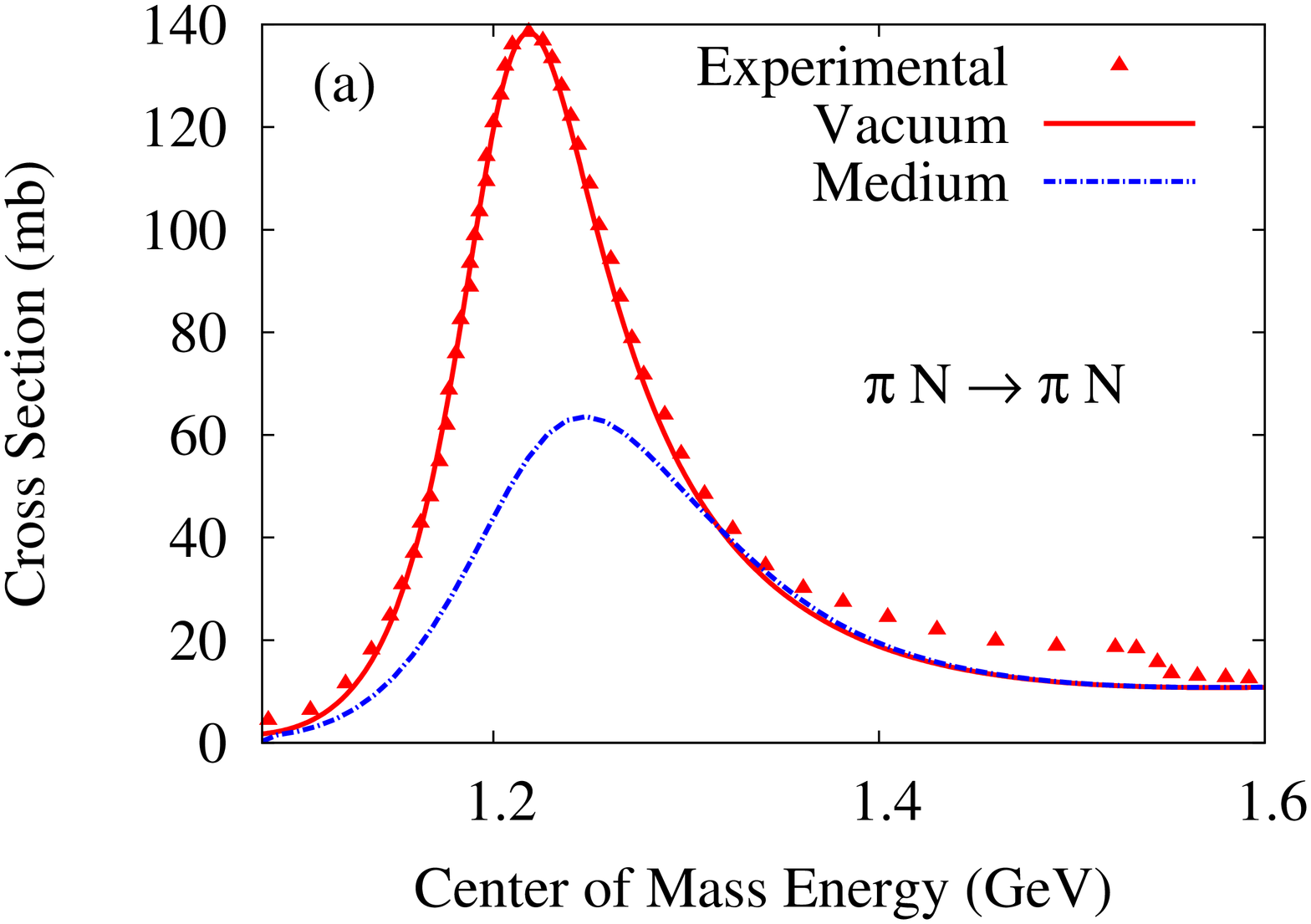} \includegraphics[scale=0.3,angle=-90]{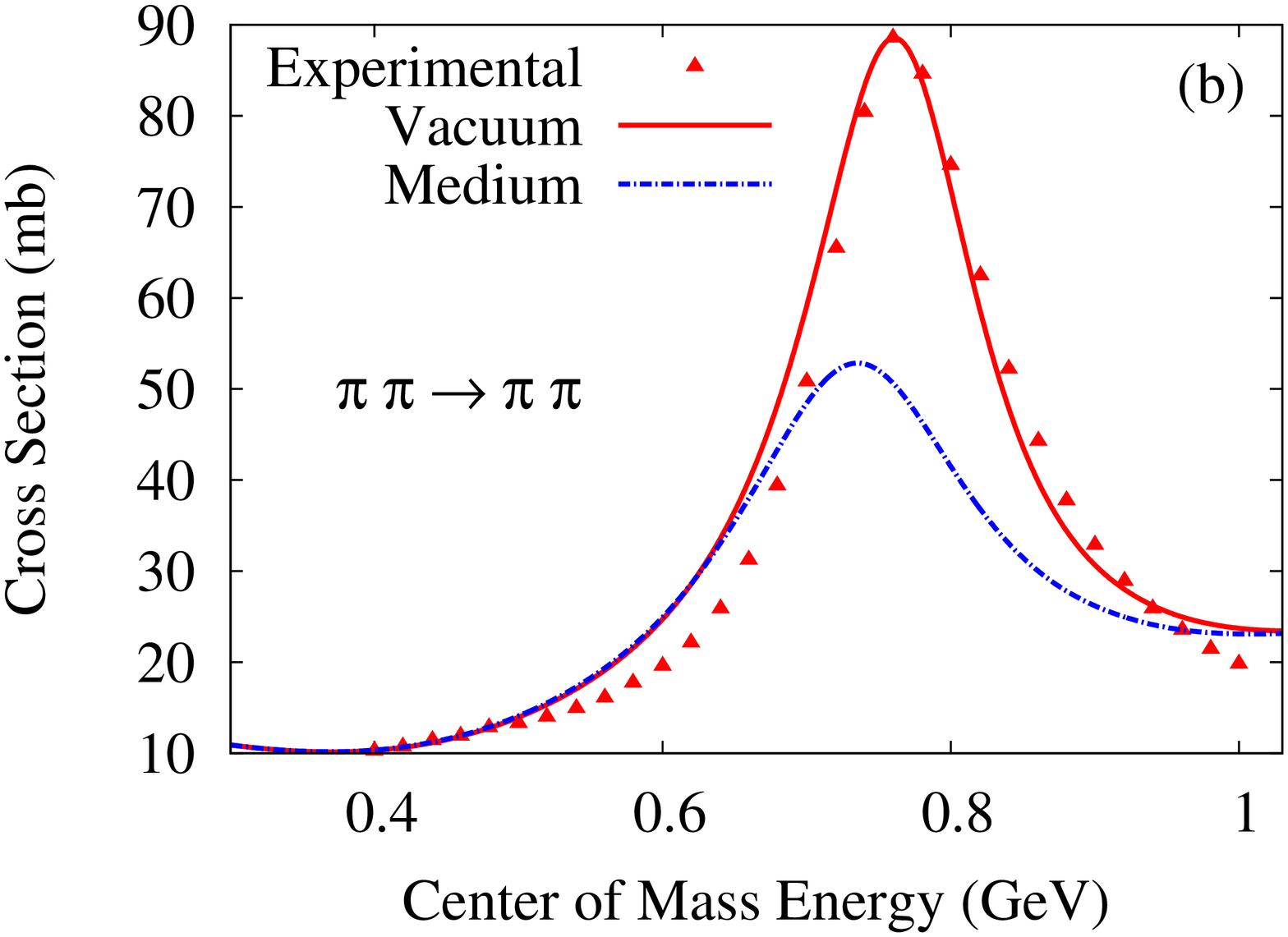}
	\caption{Isospin averaged vacuum and in-medium cross sections for 
		(a) $\pi N \rightarrow \pi N$ and (b) $\pi\pi \rightarrow \pi\pi$ scattering. Medium cross sections are obtained using 
		$T$ = 150 MeV, $\mu_\pi$ = 80 MeV and $\mu_N$ = 500 MeV. Experimental data are taken from \cite{Prakash,Bertsch:1987ux}.}
	\label{fig:xsection}
\end{figure}

In Fig.~\ref{fig:decay_width} $\Gamma_\Delta$ is plotted as a function of temperature for $\mu_\pi$ = 80 MeV and $\mu_N$ = 500 MeV. 
$\Gamma_\Delta$ increases with the increase of $T$ as can be understood from Eq.~(\ref{eq.impi.delta}) where the imaginary part 
of the self energy increases due to the increase of thermal distribution functions with the increase of $T$. This physically 
means enhancement of scattering and decay processes~\cite{HadronBook}. The effective propagator of the $\Delta$ obtained using Eqs.~(\ref{eq.repi.delta}) and (\ref{eq.impi.delta}) is used in the $s$-channel diagrams for $\pi N \rightarrow \pi N$ to arrive at a form
\begin{eqnarray} 
\overline{|\cm |^2}=\frac{1}{3}\left(\frac{f_{\pi N\Delta}}{m_\pi}\right)^2\left[\frac{A}{|s-m_\Delta^2-\Pi_\Delta |^2}+\frac{B}{(u-m_\Delta^2)^2}+\frac{C}{(u-m_\Delta^2)|s-m_\Delta^2-\Pi_\Delta |}\right]
\end{eqnarray}
where the factors $A$, $B$ and $C$ contain trace over Dirac matrices, vertex form factors etc. and can be read off from~\cite{Snigdha}. The cross-section is then given by $\sigma(s)=\frac{1}{64\pi^2 s}\int\ d\Omega\overline{|\cm |^2}$. The 
significant broadening of the width in the medium is reflected in the suppression of the cross section as can be seen 
in Fig.~\ref{fig:xsection}(a). It is to be noted that, 
the vacuum cross section (obtained using the vacuum width of the $\Delta$ in the propagator) is in very good agreement with 
the experimental data.

The $\pi\pi \rightarrow \pi\pi$ cross sections in the medium is obtained analogously. In this case, the $\pi\pi$ scattering is 
assumed to be proceed via $\rho$ and $\sigma$ meson exchange. Using $\cl=g_\rho\vec\rho^\mu\cdot\vec \pi\times\del_\mu\vec \pi+\frac{1}{2}g_\sigma m_\sigma\vec \pi\cdot\vec\pi\sigma$ with $g_\rho=6.05$ and $g_\sigma=2.5$ ~\cite{Serot:1984ey}, the spin averaged self energies 
of $\rho$ and $\sigma$ are given by,
\begin{eqnarray}
\text{Re}~\Pi_{\rho,\sigma}(q) &=& \sum\limits_{i} 
\int\frac{d^3k_\pi}{(2\pi)^3}\frac{1}{2\omega_{k_\pi}\omega_{p_i}}\mathcal{P}\left[
\left(\frac{n_+^{k_\pi}\omega_{p_i} \mathcal{N}_{\pi i,\sigma}(k_\pi^0=\omega_{k_\pi})}{(q_0-\omega_{k_\pi})^2-\omega_{p_i}^2}\right) +
\left(\frac{n_-^{k_\pi}\omega_{p_i} \mathcal{N}_{\pi i,\sigma}(k_\pi^0=-\omega_{k_\pi})}{(q_0+\omega_{k_\pi})^2-\omega_{p_i}^2}\right) \right. \nn\\ 
&&\left.+\left(\frac{n_+^{p_i}\omega_{k_\pi} \mathcal{N}_{\pi i,\sigma}(k_\pi^0=q_0-\omega_{p_i})}{(q_0-\omega_{p_i})^2-\omega_{k_\pi}^2}\right) +
\left(\frac{n_-^{p_i}\omega_{k_\pi} \mathcal{N}_{\pi i,\sigma}(k_\pi^0=q_0+\omega_{p_i})}{(q_0+\omega_{p_i})^2-\omega_{k_\pi}^2}\right) \right] \label{eq.repi.rhosigma}
\end{eqnarray}
and
\begin{eqnarray}
\text{Im}~\Pi_{\rho,\sigma}(q) &=& -\pi\epsilon(q_0)\sum\limits_{i}
\int\frac{d^3k_\pi}{(2\pi)^3}\frac{1}{4\omega_{k_\pi}\omega_{p_i}} \times \\
&& \left[\frac{}{}\mathcal{N}_{\pi i,\sigma}(k_\pi^0=\omega_{k_\pi})\left\{(1+n_+^{k_\pi}+n_+^{p_i})\delta(q_0-\omega_{k_\pi}-\omega_{p_i})
+(-n_+^{k_\pi}+n_-^{p_i}) \delta(q_0-\omega_{k_\pi}+\omega_{p_i})\right\} \right. \nn  \\
&& \left. + \mathcal{N}_{\pi i,\sigma}(k_\pi^0=-\omega_{k_\pi})\left\{(-1-n_-^{k_\pi}-n_-^{p_i})\delta(q_0+\omega_{k_\pi}+\omega_{p_i})
 + (n_-^{k_\pi}-n_+^{p_i})\delta(q_0+\omega_{k_\pi}-\omega_{p_i})\right\} \frac{}{} \right], \label{eq.impi.rhosigma}
\end{eqnarray}
where $ i \in \{\pi,\omega,h_1,a_1\}$ for the $\rho$ self energy and $i =\pi $ for $\sigma$ self energy. 
The expressions for $\mathcal{N}_{\pi i,\sigma}$ can be found in Ref.~\cite{Ghosh:2009bt}. The in-medium decay width of $\rho$ 
meson, given by,
\begin{eqnarray}
\Gamma_\rho \left( T,\mu_\pi \right) = - \frac{\text{Im}~\Pi_\rho(q^0=m_\rho,\vec{q}=\vec{0})}{m_\rho}~,
\end{eqnarray}
has been plotted against $T$ at $\mu_\pi$ = 80 MeV in Fig.~\ref{fig:decay_width}, which shows similar behaviour as $\Gamma_\Delta$ 
implying an enhancement of decay and scattering of $\rho$ with the increase of temperature. 
The in-medium $\pi\pi$ cross section obtained from the matrix elements of $\pi\pi$ scattering using the medium modified $\rho$ and $\sigma$ propagators~\cite{Sukanya2} is 
shown in Fig.~\ref{fig:xsection}(b). The vacuum cross section being in agreement with the experimental data, we find 
significant suppression in cross section at high temperature due to the broadening of $\rho$ and $\sigma$ width.


\section{Results}

\begin{figure}[h]
\includegraphics[scale=0.30,angle=-90]{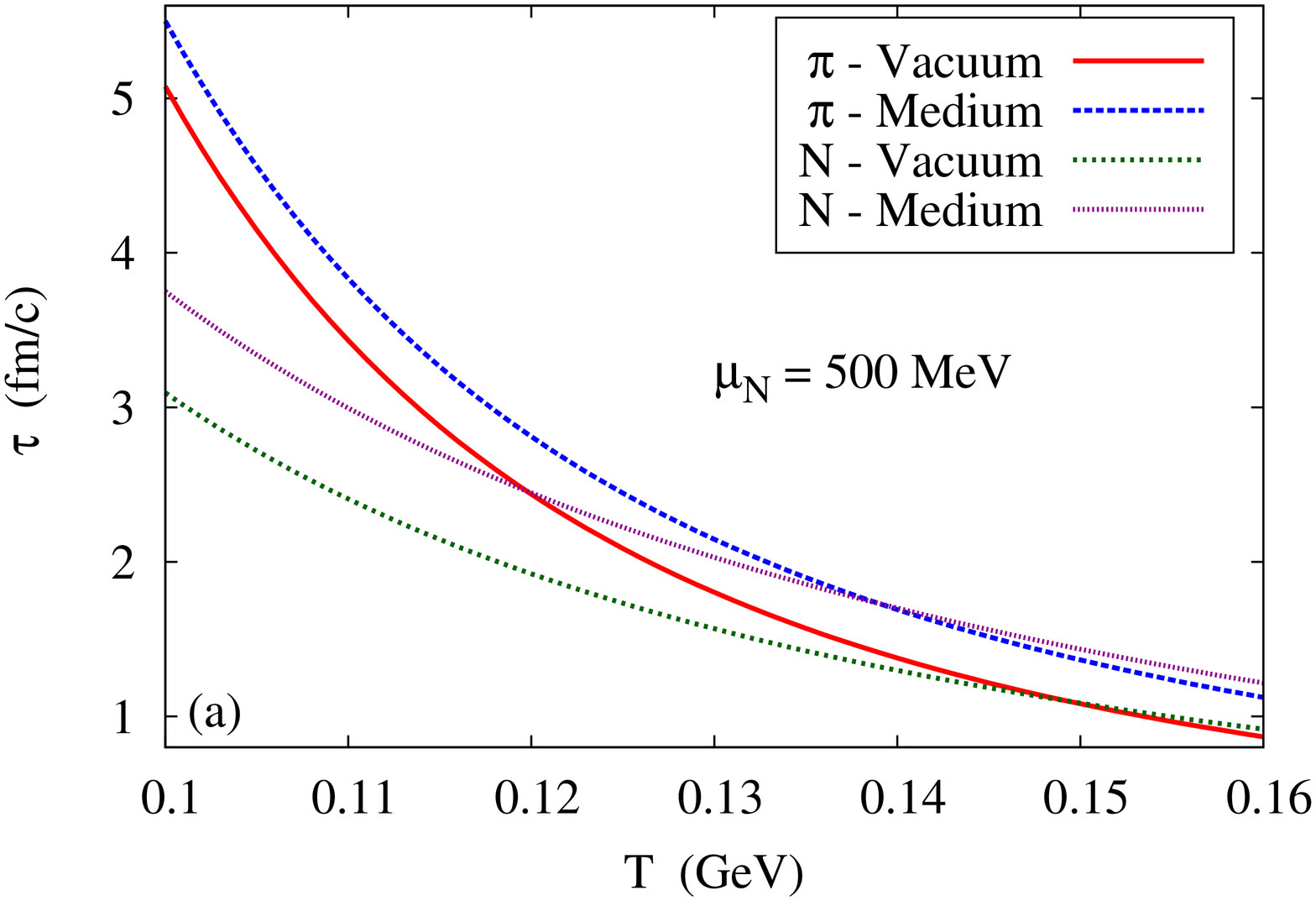} \includegraphics[scale=0.30,angle=-90]{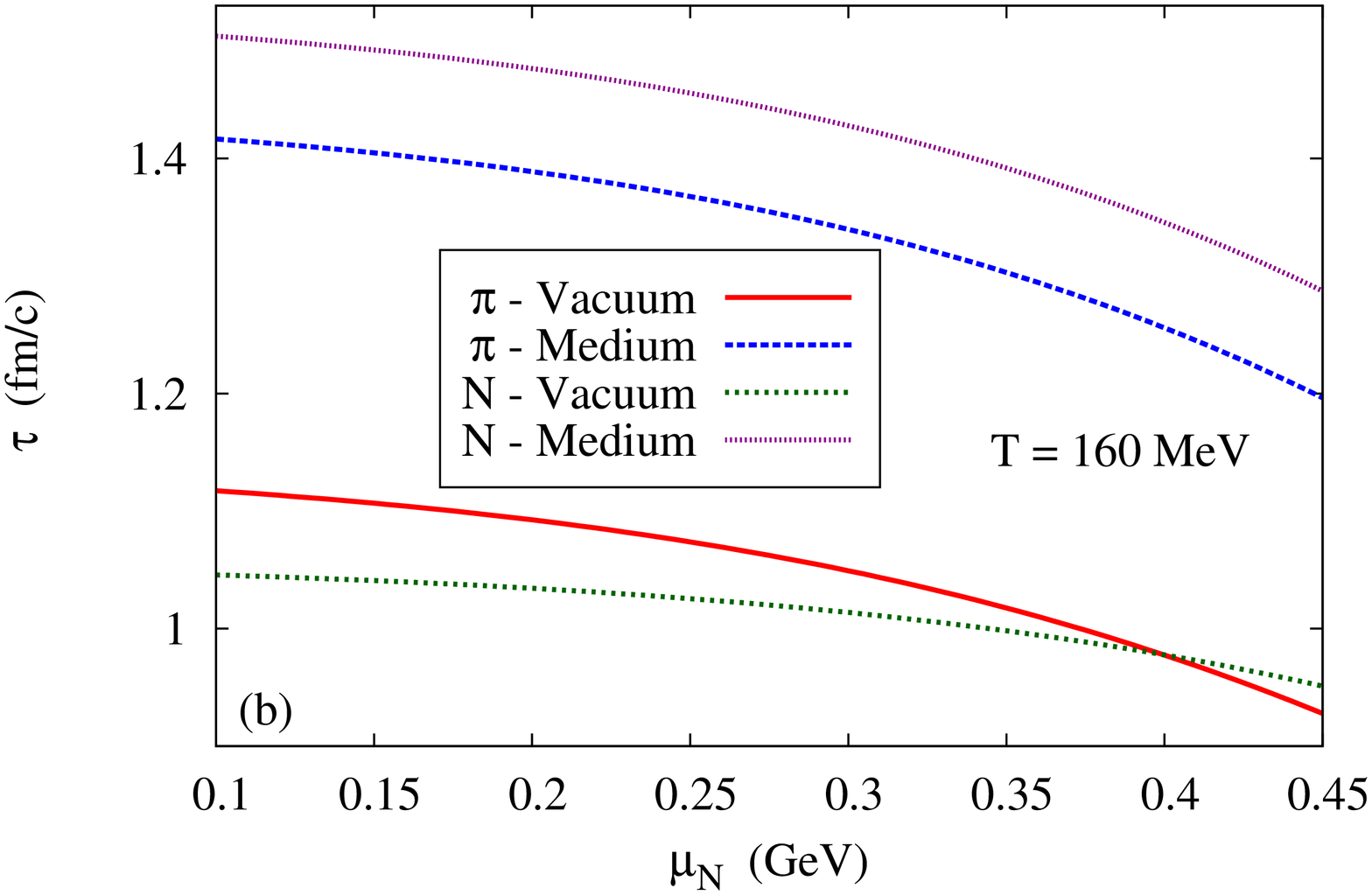}
\caption{The average relaxation time of pions and nucleons as a function of (a) $T$ for $\mu_N=500$ MeV and $\mu_\pi=80$ MeV 
		(b) $\mu$ for $T=160$ MeV and $\mu_\pi=80$ MeV}
\label{fig.tau}
\end{figure}
We first discuss the behaviour of the relaxation times of pions and nucleons making up the hadron gas mixture as a function 
of temperature and density. 
We will show numerical results for the mean relaxation time $\langle\tau_k\rangle$ corresponding to 
the $k^\text{th}$ species, which is calculated by taking the momentum average of the original $\tau_k^{-1}(p_k)$ as
\begin{eqnarray}
\langle\tau_k\rangle = \int d^3p_k f_k^{(0)} \Big/ \int d^3p_k \tau_k^{-1}(p_k) f_k^{(0)}~.
\end{eqnarray}
These contain the dynamical information of the medium embedded in the binary collisions leading to various transport 
phenomena. Both $\pi\pi$ and $\pi N$ cross-sections contribute in determining the individual relaxation times of the 
pion and nucleon as seen from Eq.~(\ref{rel_piN}). 
In Figs.~\ref{fig.tau}(a) and (b), respectively the temperature and density dependence of the pion and nucleon 
relaxation times are shown. 
The qualitative behaviour of the relaxation times with the variation of temperature and density can be understood by considering the 
(approximate) dependence of the $\tau$ on cross section ($\sigma$) and density ($n$) as $\tau\sim 1/n\sigma $; 
which for the our case of a $\pi-N$ gas mixture becomes~\cite{Itakura}
\begin{eqnarray}
\tau_\pi \simeq \frac{1}{n_\pi\sigma_{\pi \pi}+n_N\sigma_{ \pi N}r_\pi} \label{eq.taupi}\\
\tau_N \simeq \frac{1}{n_N\sigma_{NN}+n_\pi\sigma_{ N\pi }r_N} \label{eq.tauN}
\end{eqnarray}
with, $r_\pi=\frac{1}{\sqrt{2}}\sqrt{1+\frac{m_\pi}{m_N}}$ and $r_N=\frac{1}{\sqrt{2}}\sqrt{1+\frac{m_N}{m_\pi}}$. 
Thus, with the increase in $T$ or $\mu_N$, the $n_\pi$ and $n_N$ in the denominators of the above equations increase; which in turn 
causes the mean relaxation time to decreases with the increase in $T$ or $\mu_N$. Further, for a particular value of $T$, 
$n_\pi \gg n_N$ for the typical values of $\mu_N$ considering the nucleon number being Boltzmann suppressed due to its large mass. The 
factor $r_\pi \ll r_N$ implies that, the dominant contribution to $\tau_\pi (\tau_N)$ comes from the first (second) term of the 
denominator of Eqs.~\eqref{eq.taupi} and \eqref{eq.tauN}. This in turn explains the weaker $\mu_N$ dependence of the relaxation times 
as shown in Fig.~\ref{fig.tau}(b) since only the $n_N$ contains a strong $\mu_N$ dependence.

The mean relaxation time calculated using the vacuum cross-section appears to be of similar order of magnitude as in Ref.~\cite{Prakash}. 
As can be seen from the figures, $\langle\tau\rangle$ calculated using the vacuum cross section is less as compared to the same calculated 
using the in-medium cross section for a particular value of $T$ and $\mu_N$. This is due to the fact that, 
both the $\sigma_{\pi \pi}$ and $\sigma_{ \pi N}$ decrease in the medium as discussed in the previous section which in turn reduce the 
denominators of Eqs.~\eqref{eq.taupi} and \eqref{eq.tauN}. Thus, $\langle\tau\rangle$ gets enhanced when the in-medium cross sections are 
used in its calculation as compared to the same calculated using the vacuum cross sections. 
Though at lower temperatures near kinetic freeze-out the magnitude of the nucleon relaxation 
time is about a factor of two lower than that of the pions, they even out at higher temperatures. 
%
%
\begin{figure}[h]
	\includegraphics[scale=0.30,angle=-90]{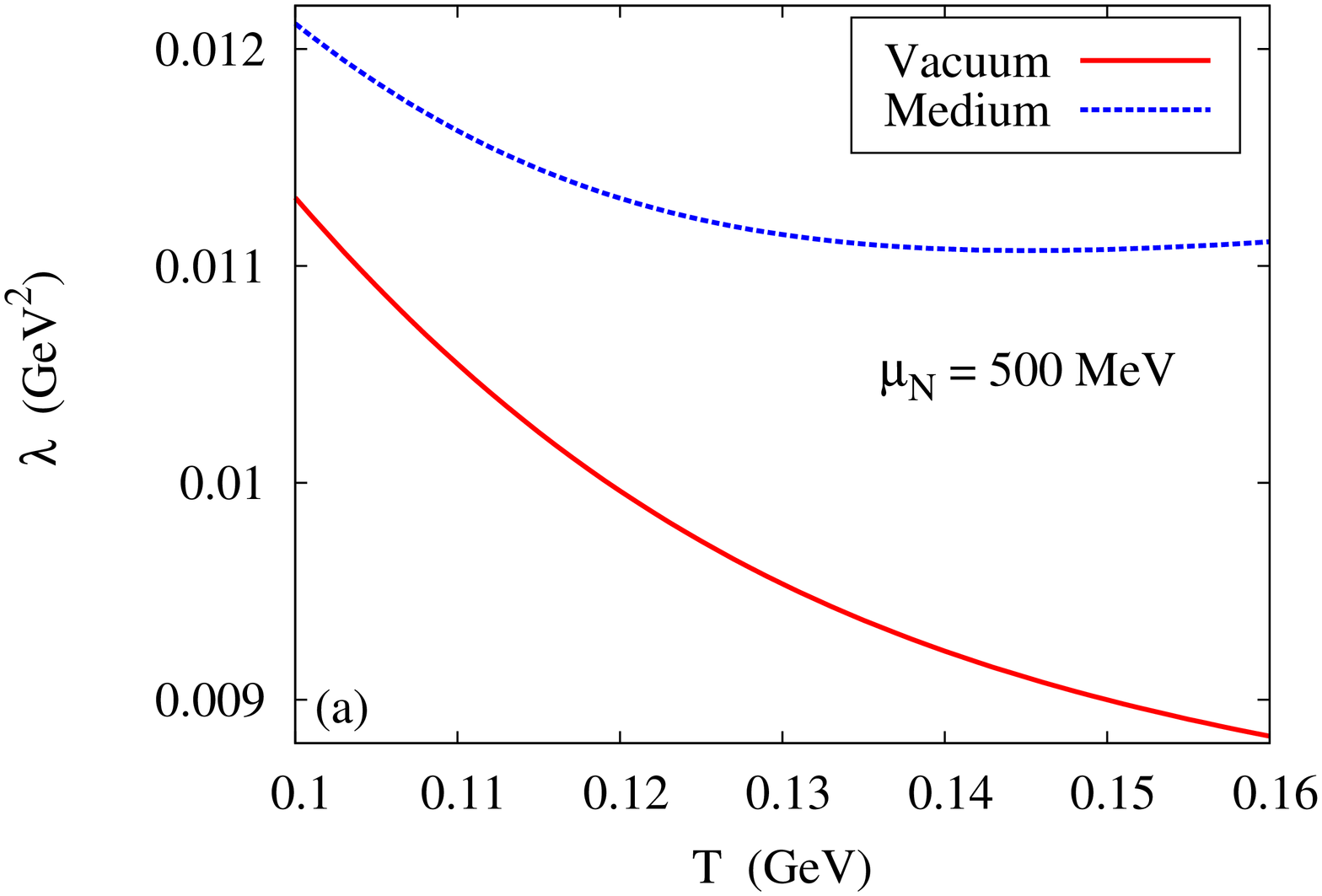} \includegraphics[scale=0.30,angle=-90]{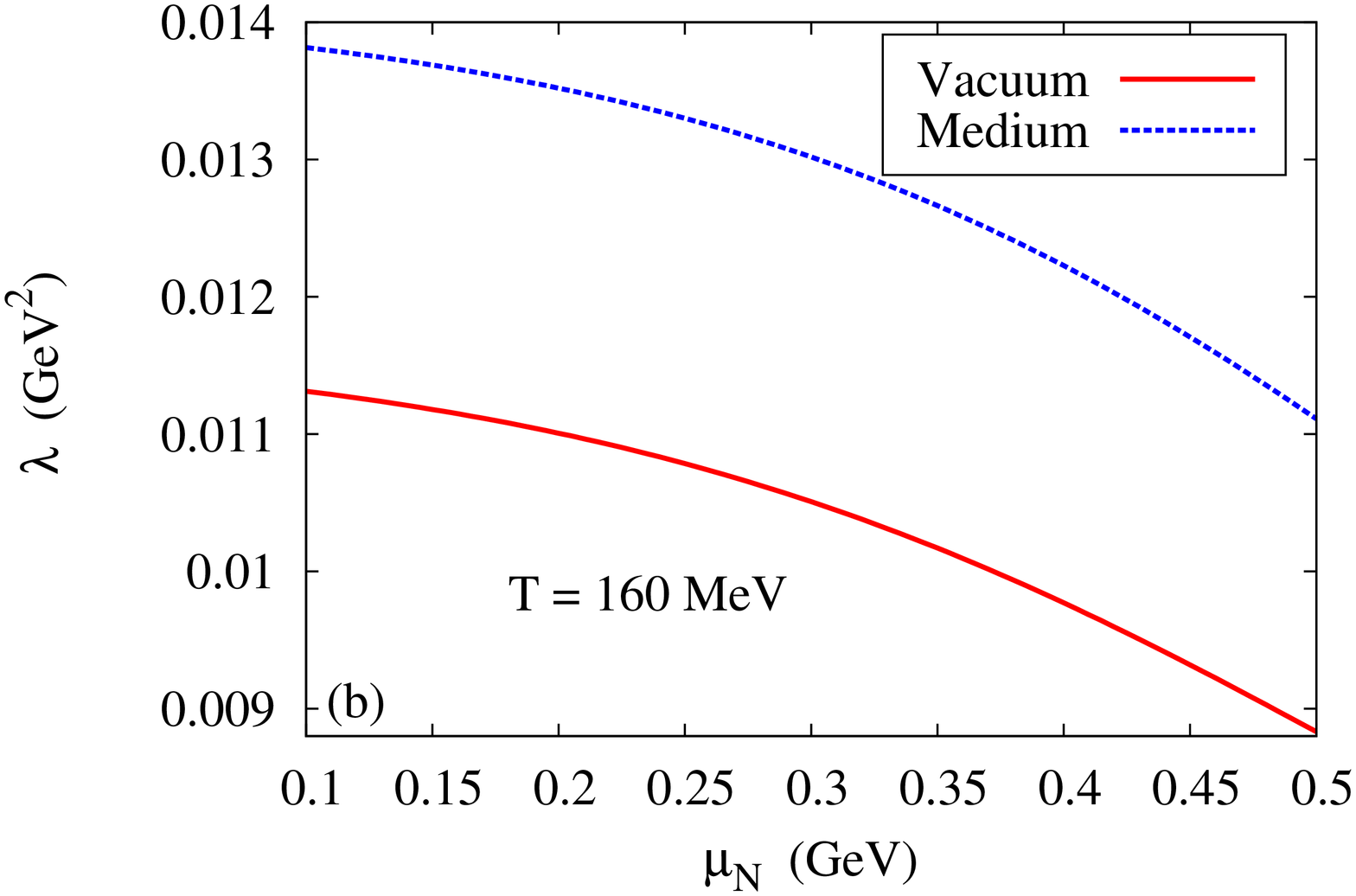}
	\caption{The thermal conductivity for a mixture constituting of nucleons and pions as 
		a function of (a) temperature and (b) nucleon density}.
	\label{fig.lambda}
\end{figure}
\begin{figure}[h]
	\includegraphics[scale=0.30,angle=-90]{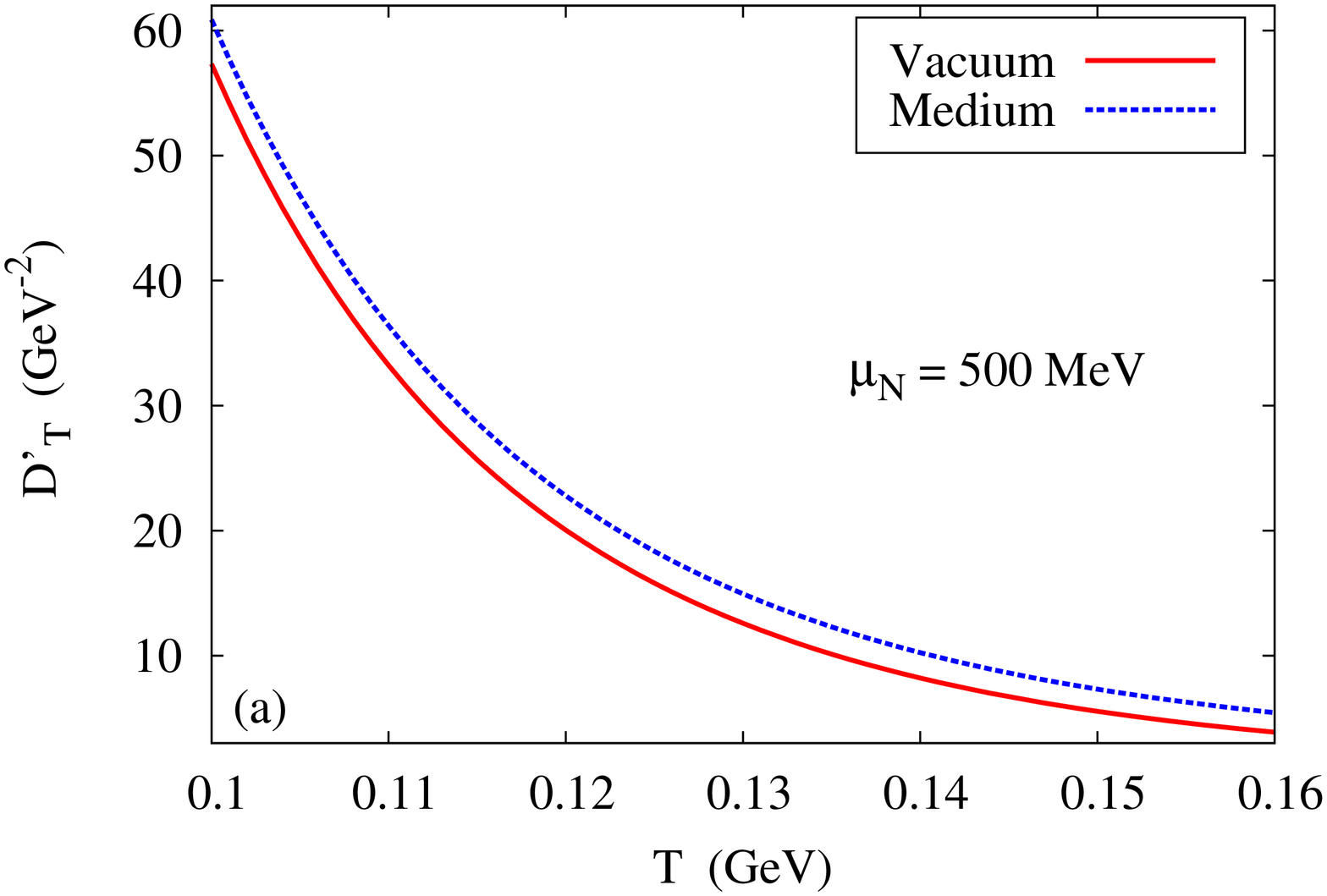} \includegraphics[scale=0.30,angle=-90]{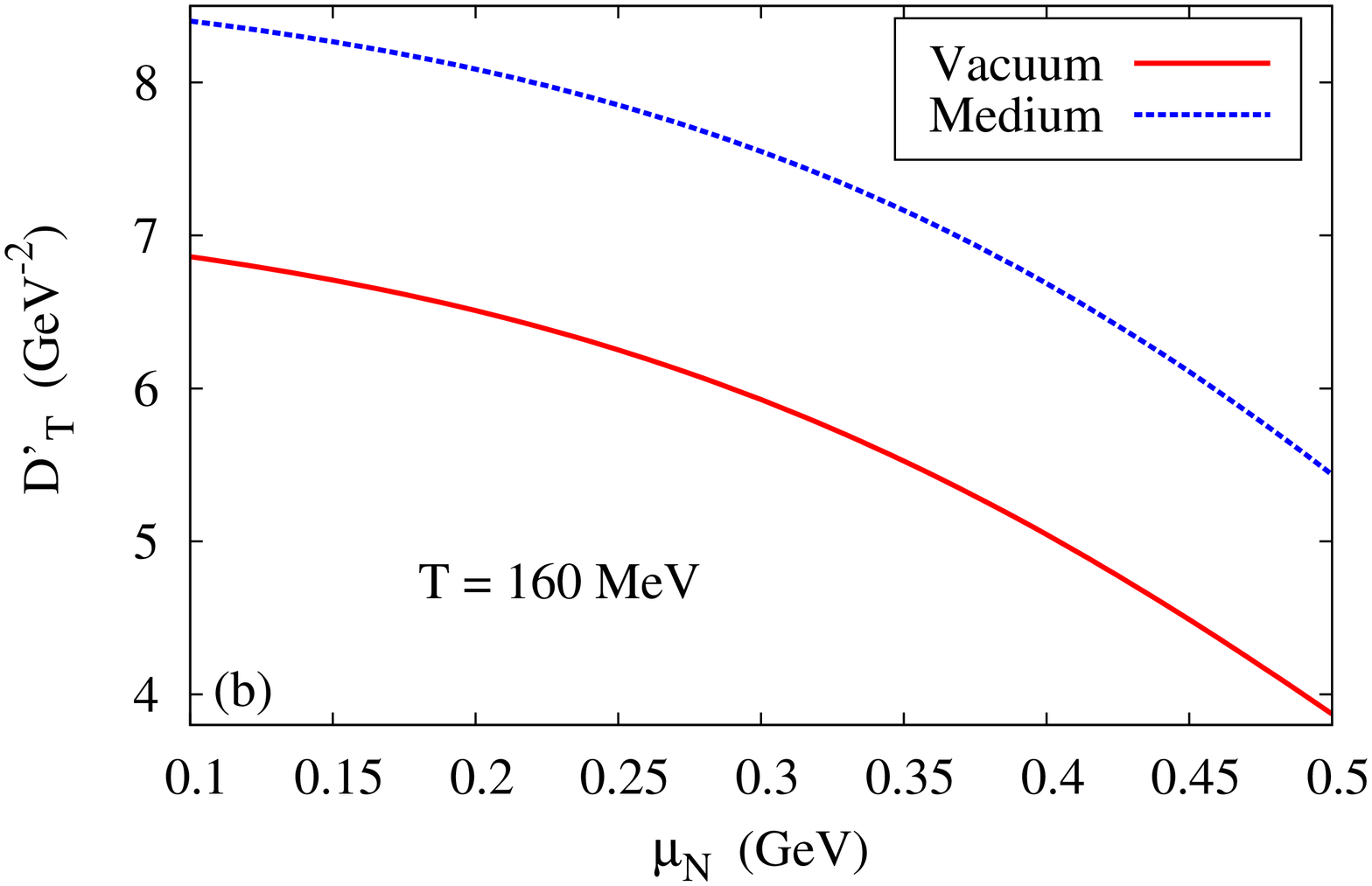}
	\caption{The Dufour coefficient as 
		a function of (a) temperature and (b) nucleon density}.
	\label{fig.Dufour}
\end{figure}
\begin{figure}[h]
	\includegraphics[scale=0.30,angle=-90]{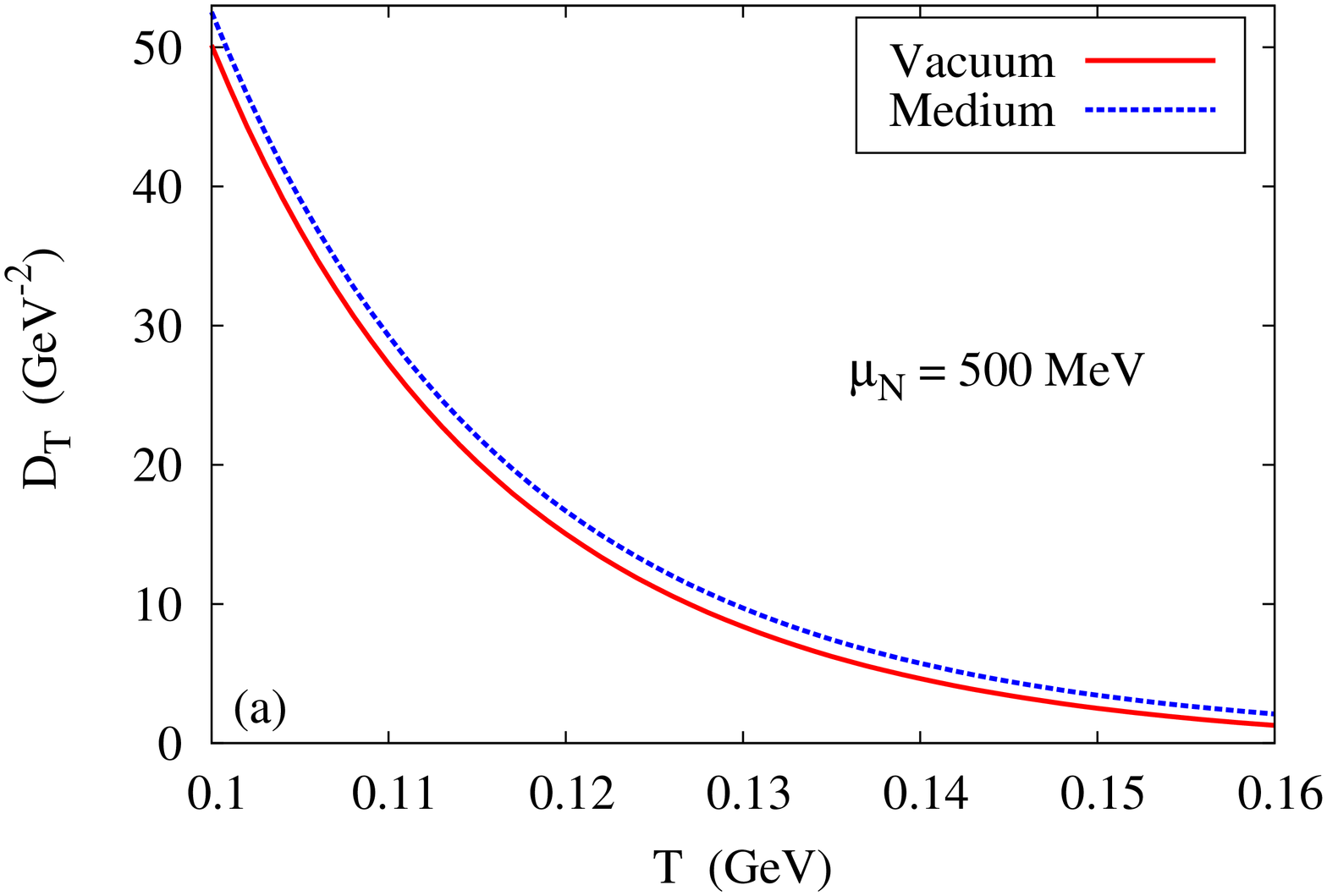} \includegraphics[scale=0.30,angle=-90]{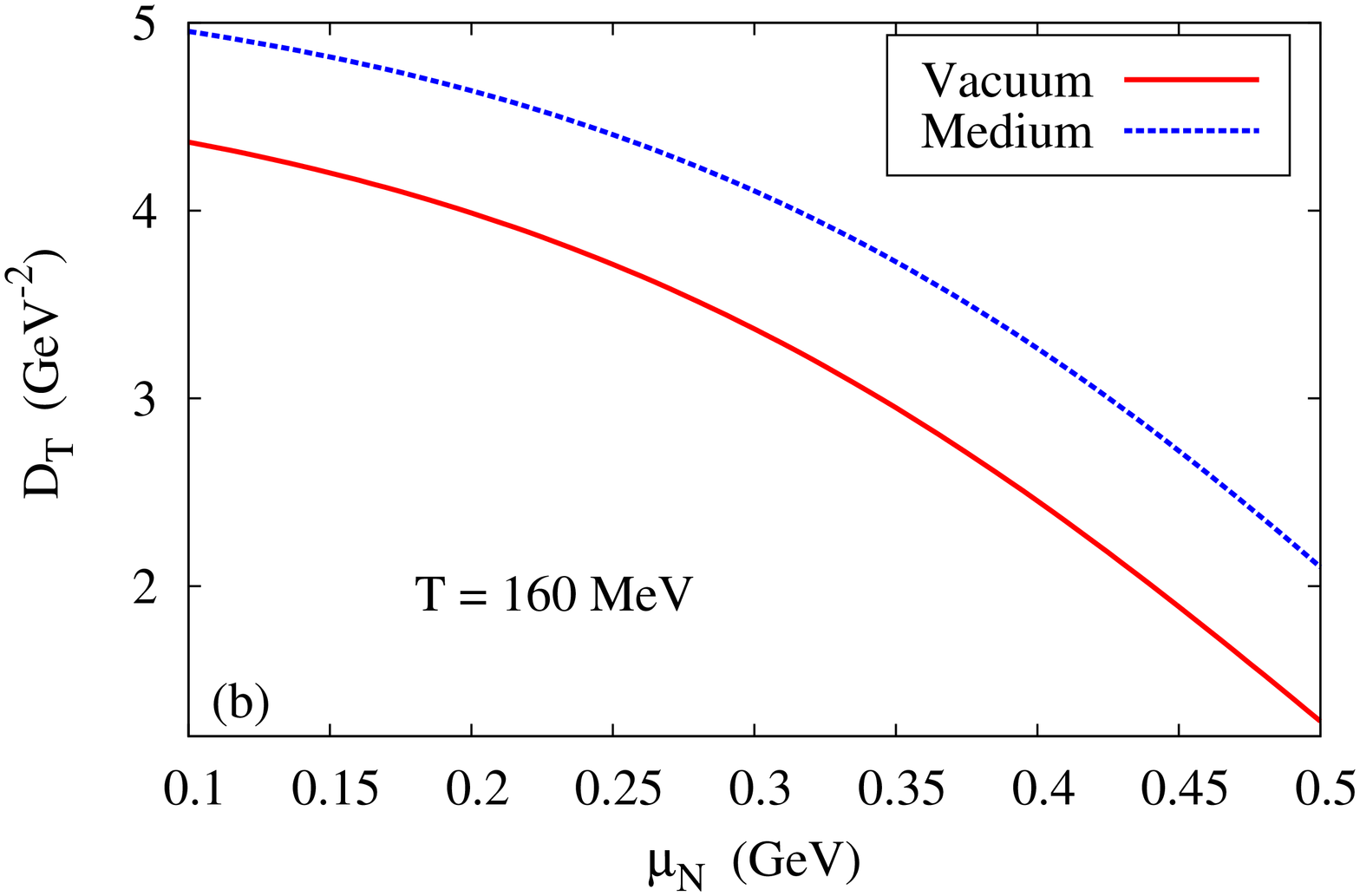} \\
	\caption{The thermal diffusion coefficient for a mixture constituting of nucleons and pions as 
		a function of (a) temperature and (b) nucleon density}.
	\label{fig.thdiff}
\end{figure}
\begin{figure}[h]
	\includegraphics[scale=0.30,angle=-90]{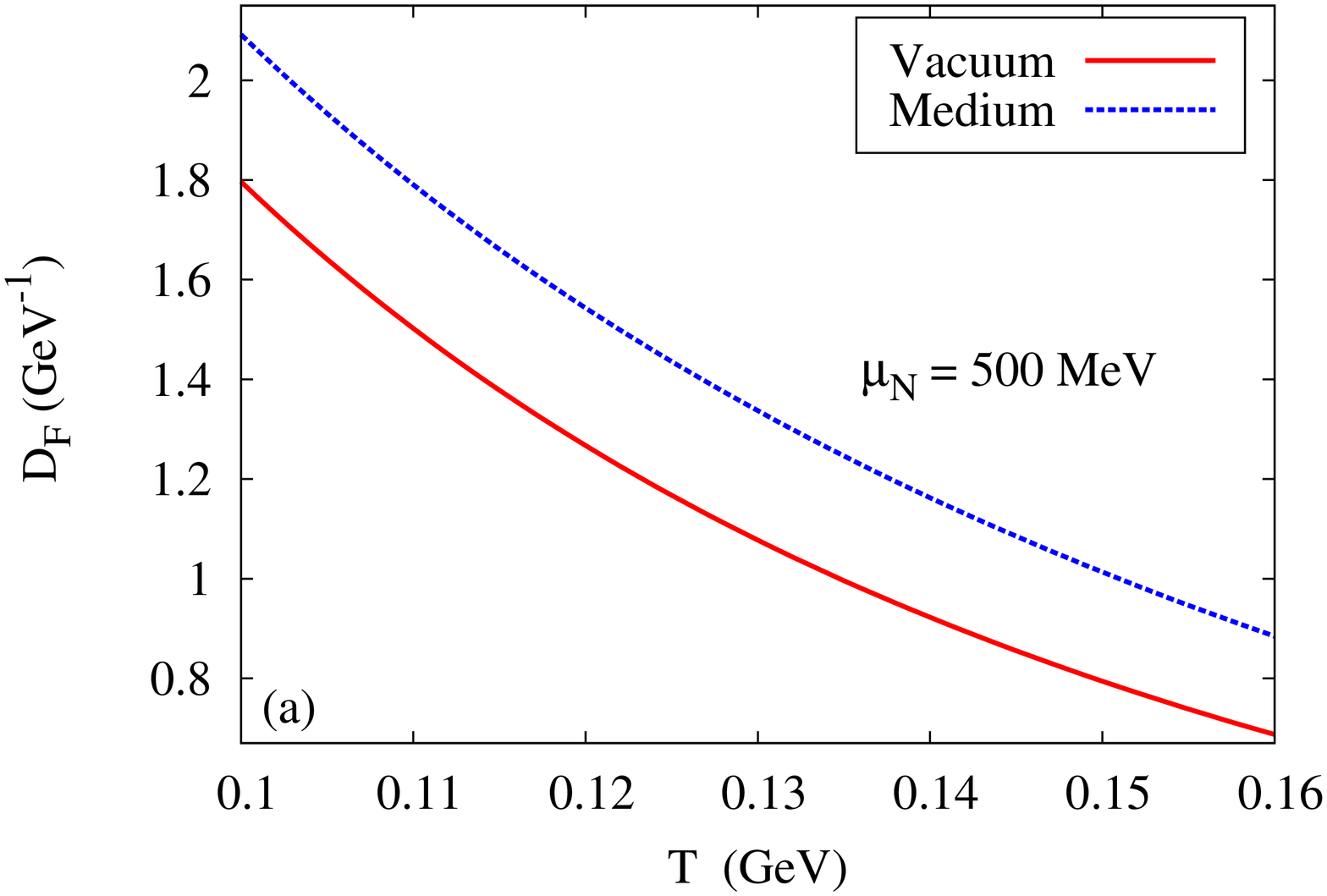} \includegraphics[scale=0.30,angle=-90]{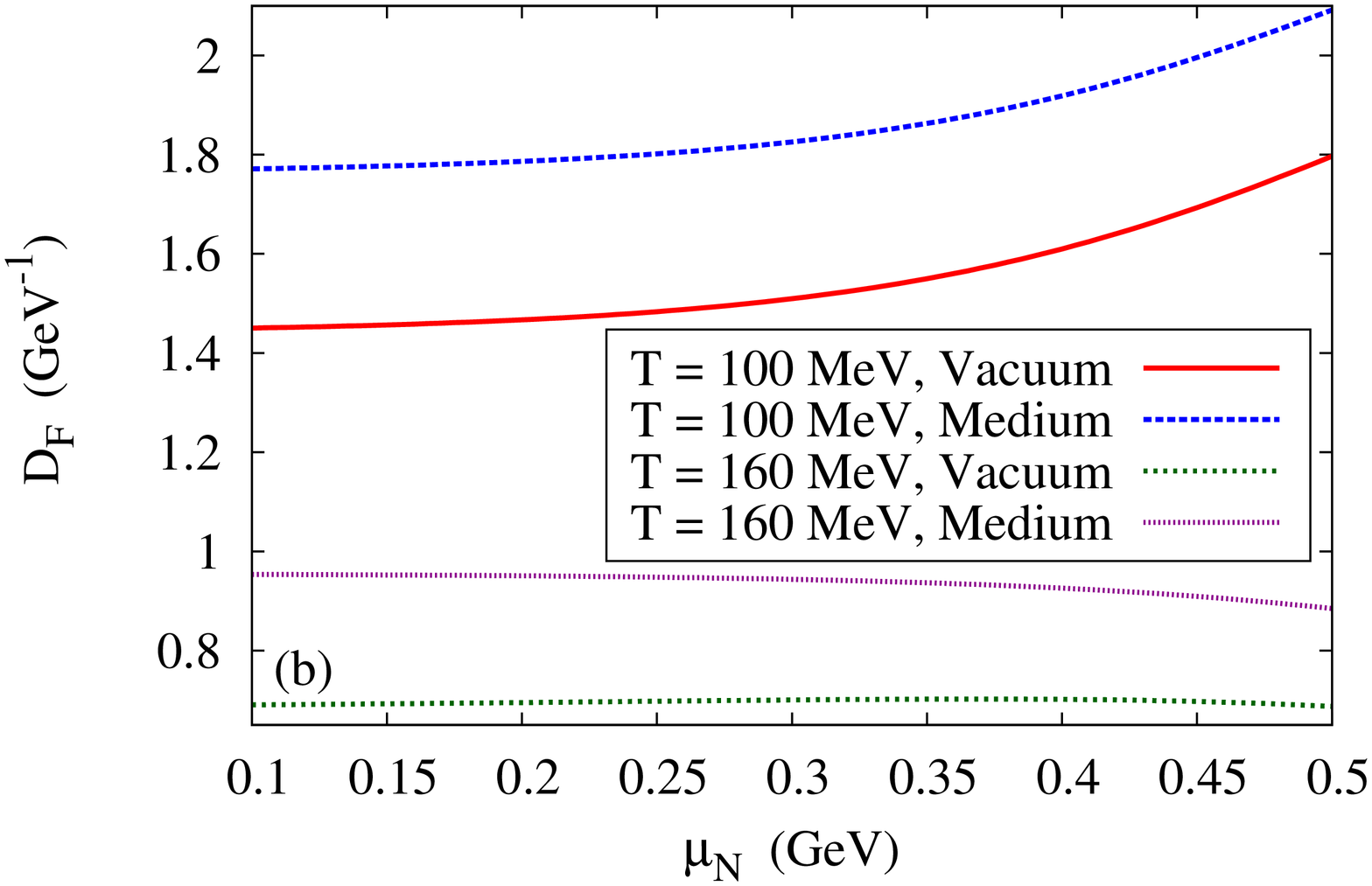} \\	
	\caption{The diffusion coefficient as 
		a function of (a) temperature and (b) nucleon density}.
	\label{fig.diffusion}
\end{figure}

We now turn to the transport coefficients as defined by Eqs.~(\ref{lambda})-(\ref{TDif}). The species $1$ and $2$ appearing in these equations denote nucleon and pion respectively. Fig.~\ref{fig.lambda}(a)  depicts the behavior of the thermal conductivity as a function of temperature where the  pion chemical potential is taken as $80$ MeV  which is its value at kinetic freezeout. The nucleon chemical potential $\mu_N$ is taken as 500 MeV. 
In order to understand the variation of $\lambda$ with the increase in temperature, let us look at Eqs.~\eqref{eq.lambda.0} 
and \eqref{eq.lambda.1} from which we notice that, $\lambda$ acquires its temperature dependence mainly through the factor 
$\sim \frac{1}{T}\cdot \tau(T) \cdot f^{(0)}(T)$. The distribution function $f^{(0)}(T)$ being an exponentially growing function 
of the temperature, the factor $\frac{1}{T} f^{(0)}(T)$ increases with the increase in $T$. On the other hand, $\tau$  
exponentially decreases as $T$ increases. Therefore, the overall behaviour of $\lambda$ with the variation of $T$ is 
determined from the competition of these two factors. As noticed in Fig.~\ref{fig.lambda}(a), when the vacuum cross 
section is used, $\lambda$ decreases almost linearly with $T$; though in the high temperature region, the slope of the curve  
slightly decreases. Also, the curve with the in-medium cross section, follows a similar trend in the low $T$ region though the 
slope of decrease is smaller than the vacuum one. However, in the high $T$ region, the $\lambda$ containing the in-medium cross section 
remains almost constant with the variation of $T$. These behaviours can be understood from the $T$ dependence of $\tau$ as 
shown in Fig.~\ref{fig.tau}(a) where, $\tau(T)$ exponentially decreases with the increase in temperature in the low temperature 
region; however in the high $T$ region, the rate of decrease becomes smaller. Moreover, the $\tau$ calculated using in-medium 
cross section has smaller slope than the same calculated using the vacuum cross section. Hence, in the high $T$ region, the two factors 
$\frac{1}{T} f^{(0)}(T)$ and $\tau(T)$ compensate each other resulting into an almost temperature independent behaviour of $\lambda$ with 
the variation of $T$ in the high temperature region.


Fig.~\ref{fig.lambda}(b) shows the variation of thermal conductivity with the nucleonic chemical potential at temperature 160 MeV. Here too we find a significant change in its value with the introduction of medium effect. The increase in average energy with temperature is canceled by the term $1/T$ in the expression of $\lambda$ and hence the thermal conductivity follows the trait of relaxation time.  Increase in density (chemical potential) is found to have very little effect on its behavior.

The variation of the Dufour and the Thermal diffusion coefficient with temperature,are presented in Fig.~\ref{fig.Dufour}(a) and Fig.~\ref{fig.thdiff}(a) respectively. Here we find that the two coefficients have roughly identical values (in fact, the values would have been exactly identical if we had used Chapman-Enskog formalism to derive the value of these coefficients). 
The concentration $x_{\pi}$ and $x_{N}$ vary very slightly with temperature, while the total concentration of particles $n$ vary as $\sim T^{3}$. Thus the density of the particles along with $T$ in the denominator plays the governing factor in determining the behavior of the Dufour and the thermal conductivity coefficients. 
Fig.~\ref{fig.Dufour}(b) and Fig.~\ref{fig.thdiff}(b) show the behavior of the Dufour coefficient and the thermal diffusion as a function of the nucleon chemical potential at $T=$160 MeV. The trend follows that of the relaxation time.
  
In Fig.~\ref{fig.diffusion}(a) we see that the diffusion coefficient decreases with with temperature. 
 On numerically analyzing the expression for the diffusion coefficient in the temperature range we have taken, it is found that, the term $L_{11}$ increases with temperature roughly as $\sim T^2$. The $n_\pi$ factor in the denominator of Eq.~\eqref{Diffusion0} is canceled by the numerator in the expression for $\Big(\frac{\partial \mu_N}{\partial x_N}\Big)_{P,T}$ in Eq.~\eqref{mu_xn}. Thus we are left with the term $2\pi^2L_{11}$ devided by the denominator of the expression in Eq.~\eqref{mu_xn}, and since the denominator varies as $\sim T^4$, the diffusion coefficient varies as $\sim 1/T^{2}$. The introduction of medium effects in the relaxation time has expected results, the value of the coefficients increases. Its variation with the nucleon chemical potential is given in Fig.~\ref{fig.diffusion}(b). It behaves quite differently from the relaxation time of collision and its value is found to depend strongly on the density of nucleons (which goes up with the increase of nucleon chemical potential). We plot for two values of $T$ in this case to show its different  behavior at lower and higher values resulting from the interplay of the two factors. At 160 MeV temperature the value of the diffusion coefficient is found to be quite steady up to $\mu_N=$450 MeV after which it decreases slowly. At lower temperatures (100 MeV) the value of diffusion coefficient goes up with increase in chemical potential.

\section{Summary}

 In the present work we have estimated the temperature and (baryon) density dependence of the thermal conductivity in addition to associated thermal and diffusion coefficients for the case of a hot hadronic gas mixture composed of nucleons and pions in the kinetic theory approach. The transport equation is solved in the Chapman-Enskog approximation and the collision term is handled in the relaxation time approximation. In order to account for the hot and dense environment we have incorporated the in-medium cross-sections for $\pi N$ and $\pi\pi$ scattering obtained using effective Lagrangians in the framework of thermal field theory. 
 Introduction of the medium effect causes a significant change in the $T$ and $\mu_N$ dependence of the transport coefficients. 
 It is observed that, the relative change of all the transport coefficients due to medium are of comparable magnitude and are more at higher temperatures and densities. This trend follows from the similar behaviour observed in the case of relaxation times. 
 It may have observable consequences on the evolution of the late stages of relativistic heavy ion collisions when included in hydrodynamic simulations.

\section*{Acknowledgement}
Snigdha Ghosh acknowledges Center for Nuclear Theory, Variable Energy Cyclotron Centre and Department of Atomic Energy, Government of India for support.

\appendix
  
\section{Appendix-A}\label{sec.appendix.a}

 The left hand side of the linearized transport equation for each species,
 \be 
 p^{\mu}_{k}U_{\mu}Df^{(0)}_{k}+p^{\mu}_{k}\bigtriangledown _{\mu}f^{(0)}_{k}=-\frac{\delta f_k}{\tau_k}~,
 \label{apC1}
 \ee  
 is to be expressed in terms of thermodynamic forces. In order to do this the derivative on the left is expressed in terms of the derivatives of the thermodynamic parameters. It is to be noted that, $D \to \del /\del t$ and $\nabla \to \partial_{i}$, i.e. the time derivative and the space derivative respectively in the local rest frame. So the above equations is written as,
 \ba
  \frac{f_{k}^{(0)}(1\pm f_{k}^{(0)})}{TE_k} \bigg\{(p_{k}\cdot U)\left[\frac{p_{k}\cdot U}{T^2}DT+D\left(\frac{\mu _{k}}{T}\right)-\frac{p^\mu_{k}}{T}
 DU_\mu\right] \nonumber\\
 +p^\mu\left[\frac{p_{k}\cdot
 U}{T^2}\nabla_\mu T+\nabla_\mu \left(\frac{\mu _{k}}{T}\right)
 -\frac{p_{k}^\nu}{T}\nabla_\mu U_\nu\right]\bigg\}=-\frac{\delta f_k}{\tau_k}~.
 \label{apC2}
 \ea
 The thermodynamic forces does not contain terms like $DT$ and $D\left(\frac{\mu_{k}}{T}\right)$; it contains the space derivative of temperature, chemical potential and the thermodynamic velocity $U^{\mu}$. In order to express the time derivative as space derivative of the thermodynamic parameters we make use of the conservation equations (which are correct up to first order), 
\begin{eqnarray}
  \del_\mu N^\mu_{k}&=&0\nonumber\\
  Dn_{k}&=&-n_{k}\del_\mu U^\mu \nonumber \\
\del_\nu T^{\mn\, (0)}&=&0 \nonumber \\
 U_\mu\del_\nu T^{(0)\,\mn}&=&0 \nonumber\\
De &=&-\frac{P}{n}\del_\mu u^\mu \nonumber \\
\sum_{k}n_{k}De_{k}&=&-\left(\sum_{k}P_{k}\right)\del_\mu U^\mu \nonumber
\end{eqnarray}
where, $N_{k}^{\mu}=n_{k}U^{\mu}$, $T^{\mu \nu}=n\left[(e+\frac{P}{n})u^\mu u^\nu-g^\mn \frac{P}{n}\right]+T^{(1)\, \mu \nu}$ 
and $T^{(1)\, \mu \nu}=\left(I^{\mu}_{q}U^{\nu}+I^{\nu}_{q}U^{\mu}\right)+~\Pi^{\mu \nu}$. 
The quantities $I_{q}^{\mu}$ and $\Pi^{\mu \nu}$ are the heat flow and the viscous part of the energy momentum tensor respectively. Expanding the equations in terms of derivative of temperature and chemical potential over temperature we get,
 
\begin{eqnarray}
 \frac{\del n_{\pi}}{\del T}DT+\frac{\del n_{\pi}}{\del \left(\mu_{\pi}/T\right)}D\left(\frac{\mu_{\pi}}{T}\right)+0 \cdot D\left(\frac{\mu_{N}}{T}\right)=-n_{\pi} \del_{\mu}U^{\mu}~~~~~~~~~~~~~~~~\nonumber\\
 \frac{\del n_{N}}{\del T}DT+0 \cdot D\left(\frac{\mu_{\pi}}{T}\right)+\frac{\del n_{N}}{\del \left(\mu_{N}/T\right)}D\left(\frac{\mu_{N}}{T}\right)=-n_{N} \del_{\mu}U^{\mu}~~~~~~~~~~~~~~~ \label{ap5}\\
 \left[n_{\pi}\frac{\del e_{\pi}}{\del T}+n_{N}\frac{\del e_{N}}{\del T}\right]DT+n_{\pi}\frac{\del e_{\pi}}{\del \left(\mu_{\pi}/T\right)}D\left(\frac{\mu_{\pi}}{T}\right)+n_{N}\frac{\del e_{N}}{\del \left(\mu_{N}/T\right)}D\left(\frac{\mu_{N}}{T}\right)=-P\delta_{\mu}U^{\mu}\nonumber
\end{eqnarray}
 
where $P=P_{\pi}+P_{N}$ is the total pressure. The different thermodynamic quantities like energy density, pressure, entropy etc. of the system consisting of pions and nucleons are expressed as follows, 
\begin{eqnarray}
n_{\pi}&=&g_{\pi}\int d\Gamma_{\pi} E_{\pi}
f_{\pi}^{(0)}(p_{\pi})=\frac{g}{2\pi^2}z_{\pi}^2T^3S_2^1(z_{\pi}),\label{paramet1}\\
P_{\pi}&=&g_{\pi}\int
 d\Gamma_{\pi}\frac{\vp_{\pi}^2}{3}f_{\pi}^{(0)}(p_{\pi})=\frac{g}{2\pi^2}z_{\pi}^2T^4S_2^2(z_{\pi}),\label{paramet2}\\
n_{\pi}e_{\pi}&=&g_{\pi}\int d\Gamma_{\pi} E_{\pi}^2f_{\pi}^{(0)}(p_{\pi})
=\frac{g_{\pi}}{2\pi^2}z_{\pi}^2T^4[z_{\pi} S_3^1(z_{\pi})-S_2^2(z_{\pi})]\label{paramet3}\\
n_{\pi}h_{\pi}&=&n_{\pi}z_{\pi}T\frac{S_3^1(z_{\pi})}{S_2^1(z_{\pi})}~,
\label{paramet4}
\end{eqnarray}
where $z_{\pi}=m_{\pi}/T$, $z_N=m_{N}/T$, $E_\pi=\sqrt{\vec{p}_\pi^2+m_\pi^2}$ and  $f_{\pi}^{(0)}(p_{\pi})=[e^{\beta(E_\pi-\mu_{\pi})}-1]^{-1}$. 
The distribution function $f_{\pi}^0$ has been expanded using the formula $[a-1]^{-1}=\displaystyle\sum_{n=1}^\infty(a^{-1})^n$, resulting in expansion of the integrals which can be compactly expressed as $S_n^\alpha(z_{\pi})=\displaystyle\sum_{k=1}^\infty e^{k\mu_{\pi}/T} k^{-\alpha} K_n(kz_{\pi})$ , $K_n(x)$ denoting the modified Bessel function of order $n$ given by
\be
K_n(x)=\frac{2^n n!}{(2n)!\ x^n}\int_x^\infty\ d\tau
(\tau^2-x^2)^{n-\frac{1}{2}}e^{-\tau}
\ee
or
\be
K_n(x)=\frac{2^n n!(2n-1)}{(2n)!\ x^n}\int_x^\infty\ \tau d\tau
(\tau^2-x^2)^{n-\frac{3}{2}}e^{-\tau}~.
\ee
 Now for the corresponding expressions of $n_N$, $P_N$, $e_N$ etc, the $S^{\alpha}_n(z_{\pi})$ will be replaced by  $T^{\alpha}_n(z_{N})$ expressed as  $T_n^\alpha(z_{N})=\displaystyle\sum_{k=1}^\infty \left(-1\right)^{k-1}e^{k\mu_{N}/T} k^{-\alpha} K_n(kz_{N})$. Using the expression in Eq.(\ref{paramet1}) to Eq.(\ref{paramet4}) we get,
\ba 
\frac{\del e_{\pi}}{\del T}&=&4z_{\pi}\frac{S_{\pi\,3}^1}{S_{\pi\,2}^1}+
z_{\pi}\frac{S_2^2S_3^0}{(S_2^1)^2}-\frac{S_2^2}{S_2^1}+z_{\pi}^2
\left[\frac{S_2^0}{S_2^1}-\frac{S_3^1S_3^0}{(S_2^1)^2}\right]\nonumber\\
\frac{\del e_{\pi}}{\del(\mu_{\pi}/T)}&=&-T\left[1-\frac{S_2^2S_2^0}{(S_2^1)^2}\right]
+Tz_{\pi}\left[\frac{S_3^0}{S_2^1}-\frac{S_3^1S_2^0}{(S_2^1)^2}\right]~.~~~\nonumber\\
\frac{\del n_{\pi}}{\del T}&=&\frac{4\pi}{\left(2\pi\right)^{2}}T^{2}\left[-z_{\pi}^{2}S^{1}_{2}+z_{\pi}^{3}S^{0}_{3}\right]~~~~~~~~~~~~~~~~~~~~~~~\nonumber\\
\frac{\del n_{\pi}}{\del \left(\mu_{\pi}/T\right)}&=&\frac{4\pi}{\left(2\pi\right)^{3}~}z_{\pi}^{2}T^{2}S_{2}^{0}~~~~~~~~~~~~~~~~~~~~~~~~~~~~~~
\label{apC6}
\ea
Putting in Eq.(\ref{ap5}) and solving for $DT$, $D\left(\mu_{\pi}/T\right)$ and $D\left(\mu_{N}/T\right)$ we get
\footnote{For corresponding derivative of $e_{N}$ and $n_{N}$ the $z_{\pi}$ and $S^{\alpha}_{\beta}$ will be replaced by $z_{N}$ and $T^{\alpha}_{\beta}$ respectively},
\ba 
T^{-1}DT=\left(1-\gamma^{'}\right)\del_{\mu}U^{\mu}\\
TD\left(\frac{\mu_{\pi}}{T}\right)=T\left[\left(\gamma_{\pi}^{''}-1\right)\hat{h}_{\pi}-\gamma_{\pi}^{'''}\right]\\
TD\left(\frac{\mu_{N}}{T}\right)=T\left[\left(\gamma_{N}^{''}-1\right)\hat{h}_{N}-\gamma_{N}^{'''}\right]
\label{apC7}
\ea 
where,
\ba 
\gamma^{'}=\frac{1}{|A|}\left\{g_{\pi}\left[z_{\pi}^{3}\left(4S_{2}^{0}S_{3}^{1}T_{2}^{0}+S_{2}^{1}S_{3}^{0}T_{2}^{0}\right)+z_{\pi}^{4}\left(\left(S_{2}^{0}\right)^2T_{2}^{0}-\left(S_{3}^{0}\right)^2T_{2}^{0}\right)\right]\right.~~~~~~~~~~~~~~~
\nonumber\\\left.+g_N\left[z_{N}^{3}\left(4S_{2}^{0}T_{2}^{0}T_{3}^{1}+S_{2}^{0}T_{2}^{1}T_{3}^{0}\right)+z_{N}^{4}\left(S_{2}^{0}\left(T_{2}^{0}\right)^2-S_{2}^{0}\left(T_{3}^{0}\right)^2\right)\right]\right\}
\ea
 
\ba
\gamma_{\pi}^{''}=\frac{1}{|A|}\left\{g_{\pi}\left[-5z_{\pi}^{2}\left(S_{2}^{1}\right)^2T_{2}^{0}+z_{\pi}^{3}\left(3S_{2}^{0}S_3^1T_2^0+3S_2^1S_3^0T_2^0\right)+z_{\pi}^{4}\left(\left(S_2^0\right)^2T_2^0-\left(S_3^0\right)^2T_2^0\right)\right]\right.\nonumber\\
\left.+g_{N}\left[-z_{N}^{2}S_2^0\left(T_2^1\right)^2+z_{N}^3\left(3S_2^0T_2^0T_3^1+2S_2^0T_2^1T_3^0\right)
+z_{N}^4\left(S_2^0\left(T_2^0\right)^2-S_2^0\left(T_3^0\right)^2\right)\right]\right\}
\ea 
\ba
\gamma_{\pi}^{'''}=\frac{1}{|A|}\left\{g_{\pi}\left[z_{\pi}^4S_2^1S_2^0T_2^0\right]+g_N[z_{N}^3\left(4S_2^1T_2^0T_3^1+S_2^1T_2^1T_3^0\right)-z_{\pi}z_{N}^2S_{3}^{0}\left(T_2^1\right)^2\right.~~~~~~~~~~~~~~~
\nonumber\\ \left.+z_{N}^4\left(S_2^1\left(T_2^0\right)^2-S_2^1\left(T_3^0\right)^2\right)
+z_{\pi}z_N^3\left(S_3^0T_2^1T_3^0-S_3^0T_2^0T_3^1\right)]\right\}
\ea

and,
\ba
|A|=g_{\pi}\left[-z_{\pi}^2\left(S_2^1\right)^2T_2^0+z_{\pi}^3\left(3S_2^0S_3^1T_2^0+2S_2^1S_3^0T_2^0\right)+z_{\pi}^4\left(\left(S_2^0\right)^2T_2^0-\left(S_3^0\right)^2T_2^0\right)\right]\nonumber\\
+g_{N}\left[-z_{N}^2S_2^0\left(T_2^1\right)^2+z_{N}^3\left(3S_2^0T_2^0T_3^1+2S_2^0T_2^1T_3^0\right)+z_{N}^4\left(S_2^0\left(T_2^0\right)^2-S_2^0\left(T_3^0\right)^2\right)\right]~.
\ea

 The corresponding expressions of $\gamma_{N}^{''}$ and $\gamma_{N}^{'''}$ are obtained by replacing $S_{\beta}^{\alpha}$ with $T_{\beta}^{\alpha}$ and vice versa in $\gamma_{\pi}^{''}$ and $\gamma_{\pi}^{'''}$ respectively.

 \section{Appendix-B}\label{sec.appendix.b}
 The Gibbs Duhem relation given by
 \ba 
 n^{-1}\partial_{\nu}P=s^{'}\partial_{\nu}T+\sum_{k=1}^{2}x_k\partial_{\nu}\mu_{k}~
 \label{GibD}
 \ea
 where $s^{'}$  is the entropy per particle, can be written as
 \ba 
 h\Big(\frac{\partial_{\nu}T}{T}-\frac{\partial_{\nu}P}{nh}\Big)+\sum_{k=1}^{2}x_kT\partial_{\nu}\Big(\frac{\mu_k}{T}\Big)=0
 \label{GibD2}
 \ea 

 The Gibbs Duhem relation shows that for a two component system there are only three independent
 intrinsic thermodynamic parameters. In Eq.(\ref{GibD}) the derivative of pressure is written in
 terms of the derivatives of the independent thermodynamic quantities which are the  temperature and 
 chemical potentials of the two species of particles. If we now change the independent intrinsic
 quantities to temperature, pressure and the concentration of the $1^{st}$ species, then 
 Eq.(\ref{GibD2}) can be written as
 \ba 
  h\Big(\frac{\partial_{\nu}T}{T}-\frac{\partial_{\nu}P}{nh}\Big)+\sum_{k=1}^{2}x_k\bigg\{T\bigg(\frac{\partial}{\partial T}\frac{\mu_k}{T}\bigg)_{P,x_1}\bigtriangledown_{\nu}T+\Big(\frac{\partial \mu_{k}}{\partial P}\Big)_{T,x_1}\bigtriangledown_{\nu}P+\Big(\frac{\partial \mu_{k}}{\partial x_1}\Big)_{P,T}\bigtriangledown_{\nu}x_1\bigg\}=0~~.
  \label{GibD3}
 \ea 
 
Now since $P$, $T$ and $x_1$ are independent parameters the coefficients of $\bigtriangledown_\nu T$, $\bigtriangledown_\nu P$ and $\bigtriangledown_\nu x_1$ must be individually zero. Hence from the coefficient of $\bigtriangledown_\nu x_1$ we get 
\ba
\Big(\frac{\partial \mu_2}{\partial x_1}\Big)_{T,P}=\frac{1}{x_2}\Big(\frac{\partial \mu_1}{\partial x_1}\Big)_{T,P}
\label{rela}
\ea
Subtracting Eq.(\ref{GibD3}) from Eq.(\ref{teq2left}) and using Eq.(\ref{rela}) we get Eq.(\ref{teq3}).

\section{Appendix-C}\label{sec.appendix.c}
Using Eq.(\ref{paramet1})and Eq.(\ref{paramet2}) along with corresponding expression for $n_N$ and $P_n$ we can calculate the partial derivative of $x_\pi$, $x_N$, $P_\pi$ and $P_N$, with respect to the independent thermodynamic parameters; $\mu_\pi/T$, $\mu_N/T$ and $T$ directly. Now if we change the independent thermodynamic parameters to $x_N$, $T$, $P$ then
\ba 
\Big(\frac{\partial P}{\partial T}\Big)_{P,x_N}=\Big(\frac{\partial P}{\partial T}\Big)_{\frac{\mu_\pi}{T},\frac{\mu_N}{T}}+\Big(\frac{\partial P}{\partial (\mu_\pi/T)}\Big)_{\frac{\mu_N}{T},T}\Big(\frac{\partial}{\partial T}\frac{\mu_\pi}{T}\Big)_{P,x_N}+\Big(\frac{\partial P}{\partial (\mu_N/T)}\Big)_{\frac{\mu_\pi}{T},T}\Big(\frac{\partial}{\partial T}\frac{\mu_N}{T}\Big)_{P,x_N}=0~~~~\\
\Big(\frac{\partial x_N}{\partial T}\Big)_{P,x_N}=\Big(\frac{\partial x_N}{\partial T}\Big)_{\frac{\mu_\pi}{T},\frac{\mu_N}{T}}+\Big(\frac{\partial x_N}{\partial (\mu_\pi/T)}\Big)_{\frac{\mu_N}{T},T}\Big(\frac{\partial}{\partial T}\frac{\mu_\pi}{T}\Big)_{P,x_N}+\Big(\frac{\partial x_N}{\partial (\mu_N/T)}\Big)_{\frac{\mu_\pi}{T},T}\Big(\frac{\partial}{\partial T}\frac{\mu_N}{T}\Big)_{P,x_N}=0~.
\ea 

Using the above two equations we can calculate the value of $\Big(\frac{\partial}{\partial T}\frac{\mu_\pi}{T}\Big)_{P,x_N}$ and $\Big(\frac{\partial}{\partial T}\frac{\mu_N}{T}\Big)_{P,x_N}$ which turns out to be
\ba 
\Big(\frac{\partial}{\partial T}\frac{\mu_\pi}{T}\Big)_{P,x_N}=\frac{[g_Nx_\pi m_N^3T_3^0-g_\pi x_Nm_\pi^3S_3^0]
	x_N-g_Nhx_\pi m_N^2T_2^0}{T^2[g_\pi x_N^2m_\pi^2S_2^0+g_N x_\pi^2m_N^2T_2^0]}\\
\Big(\frac{\partial}{\partial T}\frac{\mu_N}{T}\Big)_{P,x_N}=\frac{[g_\pi x_N m_\pi^3S_3^0-g_N x_\pi m_N^3T_3^0]x_\pi-g_\pi hx_N m_{\pi}^2S_2^0}{T^2[g_\pi x_N^2m_\pi^2S_2^0+g_N x_\pi^2m_N^2T_2^0]}
\ea 
 Using the above equations we can calculate $\beta$. Now,
 \ba
 \Big(\frac{\partial x_\pi}{\partial \mu_\pi}\Big)_{T,\mu_N}\Big(\frac{\partial \mu_\pi}{\partial x_N}\Big)_{P,T}+\Big(\frac{\partial x_\pi}{\partial \mu_N}\Big)_{T,\mu_N}\Big(\frac{\partial \mu_N}{\partial x_N}\Big)_{P,T}=1\\
\Big( \frac{\partial P}{\partial \mu_\pi} \Big)_{T,\mu_N}\Big(\frac{\partial \mu_\pi}{\partial x_N}\Big)_{P,T}+\Big( \frac{\partial P}{\partial \mu_N}\Big)_{T,\mu_N} \Big(\frac{\partial \mu_N}{\partial x_N}\Big)_{P,T}=0~.
 \ea
Using the above two equation we get
\ba 
\Big(\frac{\partial \mu_N}{\partial x_N}\Big)_{P,T}=\frac{2\pi^2n_\pi}{[g_\pi x_N^2m_\pi^2S_2^0+g_N x_\pi^2m_N^2T_2^0]}~. \label{mu_xn}
\ea

\end{document}